\documentclass[iop]{emulateapj}
\usepackage{natbib}
\usepackage{amsmath}
\usepackage{hyperref}
\usepackage{appendix}
\newcommand{\angstrom}{\text{\normalfont\AA}}
\newcommand{\Fermi}{{\em Fermi }}
\newcommand{\AssumedDistance}{{\ensuremath{ 6.5~\mathrm{kpc}}}}
\newcommand{\EjectaMass}{{\ensuremath{ (1.8 \pm 0.6) \times 10^{-5}~M_{\odot}}}}
\newcommand{\EjectaEnergy}{{\ensuremath{ (3.8 \pm 2.0) \times 10^{45}~\rm ergs}}}
\newcommand{\DerivedDistance}{{\ensuremath{ 14.8 \pm 1.6~\rm kpc}}}
\newcommand{\FillingFactor}{{\ensuremath{ (2.1 \pm 0.7) \times 10^{-2}}}}
\newcommand{\OIIIRatio}{{\ensuremath{ 29.3 \pm 4.4}}}
\newcommand{\AssumedEjectaVelocity}{{\ensuremath{ 2600 \pm 260~\rm km~s^{-1}}}}

\newcommand{\FermiGO}{{\em Fermi Gamma Ray Space Telescope}}
\newcommand{\kms}{{\ensuremath{\text{ km s}^{-1}}}}
\usepackage{color}
\usepackage[dvipsnames]{xcolor}

\begin{document}

\title{ A Detailed Observational Analysis of V1324 Sco, the Most Gamma-Ray Luminous Classical Nova to Date}

\author{Thomas Finzell\altaffilmark{1}}
\author{Laura Chomiuk\altaffilmark{1}}
\author{Brian D. Metzger\altaffilmark{2}}
\author{Frederick M. Walter\altaffilmark{3}}
\author{Justin D. Linford\altaffilmark{4,5}}
\author{Koji Mukai\altaffilmark{6,7}}
\author{Thomas Nelson\altaffilmark{8}}
\author{Jennifer H. S. Weston\altaffilmark{9}}
\author{Yong Zheng\altaffilmark{2}}
\author{Jennifer L. Sokoloski\altaffilmark{2}}
\author{Amy Mioduszewski\altaffilmark{10}}
\author{Michael P. Rupen\altaffilmark{11}}
\author{Subo Dong\altaffilmark{12}} 
\author{Sumner Starrfield\altaffilmark{13}} 
\author{C.C. Cheung\altaffilmark{14}}
\author{Charles E. Woodward\altaffilmark{15}}
\author{Gregory B. Taylor\altaffilmark{16}}
\author{Terry Bohlsen\altaffilmark{17}}
\author{Christian Buil\altaffilmark{18}}         
\author{Jose Prieto\altaffilmark{19,20}}      
\author{R. Mark Wagner\altaffilmark{21,22}}     
\author{Thomas Bensby\altaffilmark{23}}
\author{I.A. Bond\altaffilmark{24}}
\author{T. Sumi\altaffilmark{25}}
\author{D.P. Bennett\altaffilmark{26}}
\author{F. Abe\altaffilmark{27}}
\author{N. Koshimoto\altaffilmark{28}}
\author{D. Suzuki\altaffilmark{25}}
\author{P.~J. Tristram\altaffilmark{29}}
\author{Grant W. Christie\altaffilmark{30}}
\author{Tim Natusch\altaffilmark{30}}
\author{Jennie McCormick\altaffilmark{31}}
\author{Jennifer Yee\altaffilmark{32}}
\author{Andy Gould\altaffilmark{22}}
        
\altaffiltext{1}{Department of Physics and Astronomy, Michigan State Univeristy,  567 Wilson Road,  East Lansing, MI  48824-2320, USA}
\altaffiltext{2}{Columbia Astrophysics Laboratory, Columbia University, New York, NY 10027, USA}
\altaffiltext{3}{Department of Physics and Astronomy, Stony Brook University, Stony Brook, NY 11794-3800, USA}
\altaffiltext{4}{Department of Physics, The George Washington University, Washington, DC 20052, USA}
\altaffiltext{5}{Astronomy, Physics, and Statistics Institute of Sciences, The George Washington University, Washington, DC 20052, USA}
\altaffiltext{6}{CRESST and X-ray Astrophysics Laboratory, NASA/GSFC, Greenbelt, MD 20771, USA}
\altaffiltext{7}{Department of Physics, University of Maryland, Baltimore County, 1000 Hilltop Circle, Baltimore, MD 21250, USA}
\altaffiltext{8}{School of Physics and Astronomy, University of Minnesota, 116 Church Street SE, Minneapolis, MN 55455, USA}
\altaffiltext{9}{Green Bank Observatory,P.O. Box 2, Green Bank, WV 24944 USA}
\altaffiltext{10}{National Radio Astronomy Observatory, P.O. Box O, Socorro, NM 87801,USA}
\altaffiltext{11}{National Research Council of Canada, Herzberg Astronomy and Astrophysics Programs, Dominion Radio Astrophysical Observatory, Canada}
\altaffiltext{12}{Kavli Institute for Astronomy and Astrophysics, Peking University, Yi He Yuan Road 5, Hai Dian District, Beijing 100871, China}
\altaffiltext{13}{School of Earth and Space Exploration, Arizona State University, Tempe, AZ 85287-1404, USA}
\altaffiltext{14}{Space Science Division, Naval Research Laboratory, Washington, DC 20375-5352, USA}
\altaffiltext{15}{Minnesota Institute for Astrophysics, School of Physics and Astronomy, University of Minnesota, 116 Church Street S.E., Minneapolis,MN 55455}
\altaffiltext{16}{Department of Physics and Astronomy, University of New Mexico, Albuquerque, NM, USA}
\altaffiltext{17}{Mirranook Observatory, Boorolong Rd Armidale, NSW, 2350, Australia}
\altaffiltext{18}{Castanet Tolosan Observatory, 6 place Clemence Isaure, 31320 Castanet Tolosan, France}
\altaffiltext{19}{N\'ucleo de Astronom\'ia de la Facultad de Ingenier\'ia, Universidad Diego Portales, Av. Ej\'ercito 441, Santiago, Chile}
\altaffiltext{20}{Millennium Institute of Astrophysics, Santiago, Chile} 
\altaffiltext{21}{LBT, University of Arizona, 933 N. Cherry Ave, Room 552, Tucson, AZ 85721, USA}
\altaffiltext{22}{Department of Astronomy, The Ohio State University, Columbus, OH 43210, USA}
\altaffiltext{23}{Lund Observatory, Department of Astronomy and Theoretical Physics, Box 43, SE-221 00 Lund, Sweden}
\altaffiltext{24}{Institute of Information and Mathematical Sciences, Massey University, Private Bag 102-904, North Shore Mail Centre, Auckland, New Zealand}
\altaffiltext{25}{Department of Earth and Space Science, Graduate School of Science, Osaka University, Toyonaka, Osaka 560-0043, Japan}
\altaffiltext{26}{Laboratory for Exoplanets and Stellar Astrophysics, NASA/Goddard Space Flight Center, Greenbelt, MD 20771, USA}
\altaffiltext{27}{Institute for Space-Earth Environmental Research, Nagoya University, Nagoya 464-8601, Japan}
\altaffiltext{28}{Department of Earth and Space Science, Graduate School of Science, Osaka University, Toyonaka, Osaka 560-0043, Japan}
\altaffiltext{29}{Mt. John University Observatory, P.O. Box 56, Lake Tekapo 8770, New Zealand}
\altaffiltext{30}{Auckland Observatory, Auckland, New Zealand}
\altaffiltext{31}{Farm Cove Observatory, Centre for Backyard Astrophysics, Pakuranga, Auckland, New Zealand}
\altaffiltext{32}{Harvard-Smithsonian Center for Astrophysics, 60 Garden Street, Cambridge, MA 02138 USA}

\begin{abstract}
It has recently been discovered that some, if not all, classical novae emit GeV gamma rays during outburst, but the mechanisms involved in the production of the gamma rays are still not well understood. We present here a comprehensive multi-wavelength dataset---from radio to X-rays---for the most gamma-ray luminous classical nova to-date, V1324 Sco. Using this dataset, we show that V1324 Sco is a canonical dusty Fe-II type nova, with a maximum ejecta velocity of 2600 km s$^{-1}$ and an ejecta mass of few $\times 10^{-5}$ M$_{\odot}$. There is also evidence for complex shock interactions, including a double-peaked radio light curve which shows high brightness temperatures at early times.
To explore why V1324~Sco was so gamma-ray luminous, we present a model of the nova ejecta featuring strong internal shocks, and find that higher gamma-ray luminosities result from higher ejecta velocities and/or mass-loss rates. Comparison of V1324~Sco with other gamma-ray detected novae does not show clear signatures of either, and we conclude that a larger sample of similarly well-observed novae is needed to understand the origin and variation of gamma rays in novae.

\end{abstract}


\section{Introduction}
\label{sec:intro}
Classical novae are the result of a thermonuclear runaway taking place on the surface of a white dwarf and are fueled by matter accreted onto the white dwarf from a companion star. These outbursts give rise to an increase in luminosity across the electromagnetic spectrum, and eject $\sim 10^{-3} - 10^{-7} M_{\odot}$ at velocities $>10^3$ \kms \citep{1978ARA&A..16..171G, 1986ApJ...310..222P, 2005ApJ...623..398Y, 2012BASI...40..185S,2016PASP..128e1001S}. 

\subsection{Novae at GeV Energies}
Nova outbursts have been detected in the GeV gamma-ray regime with the \FermiGO\ and its Large Area Telescope (LAT; see e.g.~\citealt{2010ATel.2487....1C, 2012ATel.4224....1C, 2012ATel.4284....1C, 2015ATel.7315....1C, 2016ApJ...826..142C, 2013ATel.5302....1H, 2014Sci...345..554A}). They show gamma-ray luminosities ($>$ 100 MeV) of $10^{34}-10^{36}$ erg s$^{-1}$, lasting for 2--8 weeks around optical maximum \citep{2014Sci...345..554A, 2016ApJ...826..142C}.

The presence of gamma rays implies that there are relativistic particles being generated in the nova event. There are two potential classes of processes for producing gamma rays from relativistic particles: leptonic and hadronic. In the leptonic class, electrons are accelerated up to relativistic speeds, and produce gamma rays via inverse Compton and/or relativistic bremmstrahlung processes~\citep{1970RvMP...42..237B, 2016arXiv161104532V}. In the hadronic process, it is ions that are being accelerated to relativistic speeds; these particles collide with a dense medium to produce $\pi^0$ mesons, which then decay to gamma rays \citep{1994A&A...287..959D}. The likely source of the accelerated particles is strong shocks, which can accelerate particles to relativistic speeds via the diffusive shock acceleration mechanism \citep{1978ApJ...221L..29B, Bell78, 2014MNRAS.442..713M}. 

The first nova detected by \Fermi/LAT was V407 Cyg, and it received considerable attention~\citep{2010Sci...329..817A, 2012ApJ...754...77A, 2012ApJ...761..173C, 2012BaltA..21...47E, 2012BaltA..21...54M, 2012ApJ...748...43N, 2012MNRAS.419.2329O, 2012A&A...540A..55S, 2013A&A...551A..37M}. Given that V407 Cyg has a Mira giant secondary with a dense wind (a member of the symbiotic class of systems), a model to explain the gamma rays was proposed wherein a shock was generated as the nova ejecta interacted with the dense ambient medium. A similar model was proposed to explain the inferred presence of relativistic particles in another star with a giant wind, RS Oph~\citep{2007ApJ...663L.101T}.

\begin{figure*}[ht]
\includegraphics[width=2.0\columnwidth]{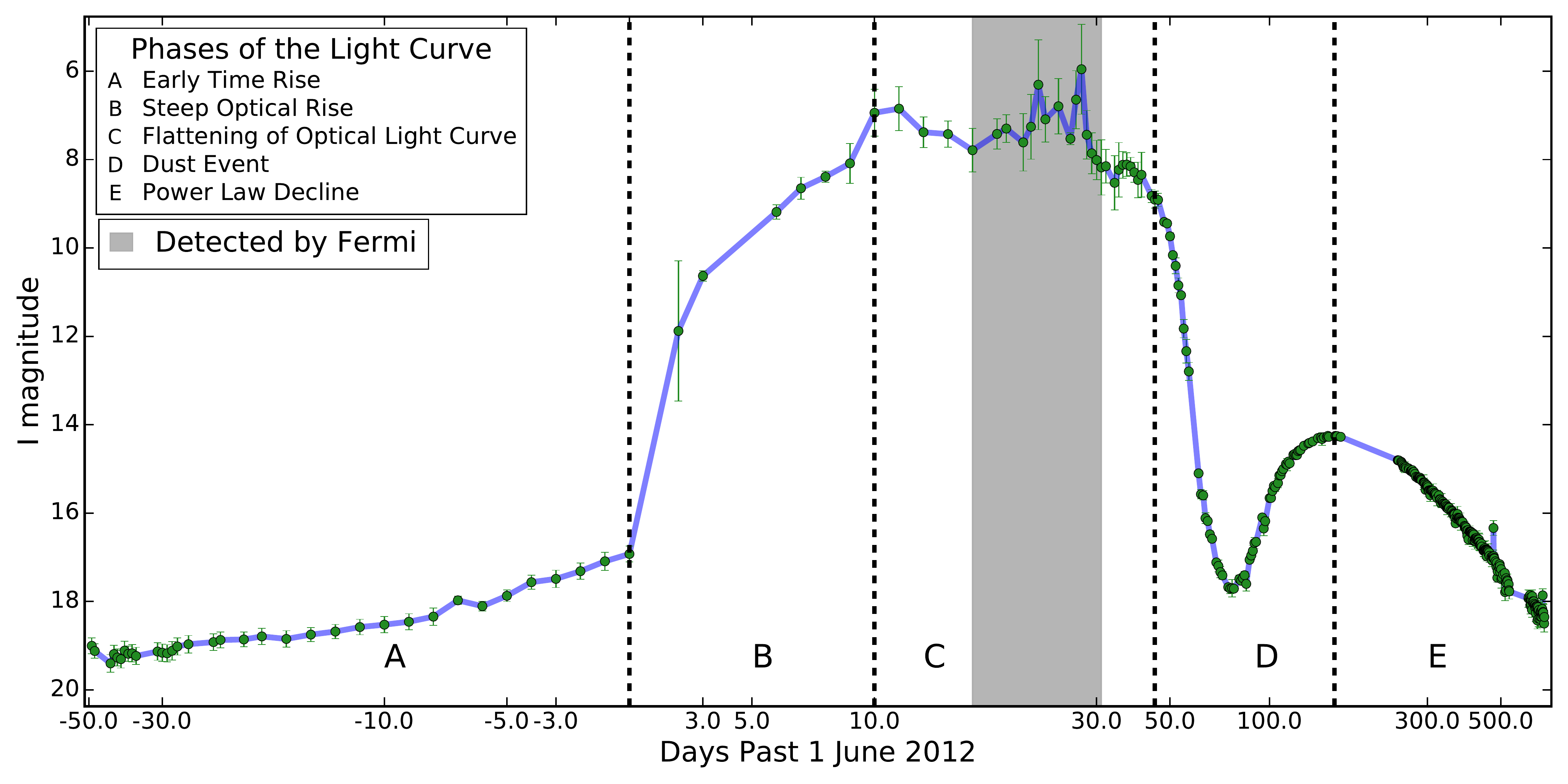}
\caption{\label{fig:OpticalLC}
$I$ band light curve for V1324 Sco, generated using the MOA data set. The plot starts 49 days before primary optical rise, at the first date where a single observation (as opposed to a stacked observation) yields a $5\sigma$ detection. The dashed lines delineate the different phases of the light curve evolution, as described in Section~\ref{sec:OpticalPhotometry}. The gray shaded region denotes the time period when V1324 Sco was detected in gamma rays. Thanks to the extremely well-sampled MOA data set, we can see all of the different evolutionary phases of the optical light curve, as discussed in section~\ref{sec:OpticalTimeline}. Note, that the X-axis takes the date of the primary optical rise (2012 June 1) to be day 0, so the plot starts on a negative value. 
}\end{figure*}

These models, however, could not explain subsequent novae detected by \emph{Fermi}/LAT: V1324 Sco, V959 Mon, V339 Del~\citep{2014Sci...345..554A} and V1369 Cen and V5668 Sgr~\citep{2016ApJ...826..142C}. These systems do not have a detectable red-giant companion (see, e.g.,~\citealt{2015ApJ...809..160F,2013MNRAS.435..771M,2013IBVS.6087....1M,2013ATel.5621....1H}; although note that the underlying binary of V5668 Sgr has yet to be identified). While it is theoretically possible for these novae to have high density circumstellar material despite not having a red-giant companion~\citep{2001ApJ...548..900S}, no evidence has yet been found for substantial circumstellar material around cataclysmic variables~\citep{2013AJ....145...19H}. Therefore, the non-detection of red-giant companions implies that these novae have main-sequence companions with low-density circumstellar material. 

It is in fact much more likely that the shocks are being produced within the ejecta, due to different components of the ejecta colliding with one another (internal shocks).  There has already been long standing evidence for internal shocks in classical novae from X-ray observations \citep[e.g.,][]{OBrien94, 2001ApJ...551.1024M}. One idea for generating internal shocks in novae was put forward by~\cite{2014Natur.514..339C}, supported by radio imaging. First, the binary interacts with the puffed-up nova envelope, resulting in a relatively slow flow with density enhancement along the binary orbital plane. Later, a separate, fast wind is launched from the white dwarf. When these two outflows collide with one another, they produce shocks. Progress has been made on the theoretical front, by constraining the shock conditions necessary to explain both the thermal and non-thermal emission observed in gamma-ray detected novae, and finding them consistent with condition expected in novae \citep{2014MNRAS.442..713M, 2016MNRAS.463..394V, 2016MNRAS.457.1786M, 2016arXiv161104532V}. 
 

\subsection{V1324~Sco}
The goal of this paper is to use a multi-wavelength analysis of the gamma-ray detected nova V1324~Sco, in order to assess the causes and energetics of the shocks that yield GeV gamma rays. V1324~Sco was originally discovered at optical wavelengths by the Microlensing Observations in Astrophysics collaboration (MOA; \citealt{2012ATel.4157....1W}). From their high-cadence monitoring of the Galactic bulge, MOA noticed a transient appear on 2012 May 22.8 UT  at RA = 17h50m53.90s and Dec = $-32^{\circ}37^{\prime}20.46^{\prime\prime}$ (J2000). Between 2012 May 22--2012 June 1, this source brightened only gradually, and its brightness was modulated with a periodicity of $\sim$1.6 hr. Starting on 2012 June 1, the source brightened much more rapidly (see Section~\ref{sec:OpticalPhotometry}), and a spectrum obtained on 2012 June 4 identified the transient as a classical nova \citep{2012ATel.4157....1W}. In retrospect, the gradual brightening over the week prior to June 1 is likely a nova ``precursor" event; similar events have been observed in a handful of other novae but remain poorly understood \citep{Collazzi09}. The precursor event is outside the scope of this paper and will be the subject of Wagner et al. (2017, in preparation), although it can be see in Figure 1.

Shortly thereafter, the transient was found to be associated with GeV gamma rays, as detected by \emph{Fermi}/LAT in its all-sky survey mode \citep{2012ATel.4284....1C}. The gamma rays were detected during the time range 2012 Jun 15--July 2 with an average flux, $\sim 5 \times 10^{-7}$ photons cm$^{-2}$ s$^{-1}$ ($>$100 MeV; \citealt{2014Sci...345..554A}). Ackermann et al.\ also point out that the spectrum of V1324~Sco may extend to higher energies than the other gamma-ray detected novae, although the spectral analysis suffers from poor statistics above a few GeV. 

Multi-wavelength observations were initiated during the period of gamma-ray detection, including observations at radio (\citealt{Chomiuk12}; Section~\ref{sec:RadioSection}), infrared \citep{Raj12}, and X-ray wavelengths (\citealt{Page12, Page13}; Section~\ref{sec:XRay}). High-resolution spectra observed during the nova outburst allowed our team to measure absorption features, enabling an estimate of reddening and placing a lower limit on the distance \citep{2015ApJ...809..160F}. We estimate $E(B-V) = 1.16 \pm 0.12$. Three-dimensional reddening maps of the Galaxy \citep{Schultheis14} imply that V1324~Sco is $>$ 6.5 kpc away (see also \citealt{Munari15} for similar results). V1324~Sco's location near or beyond the Galactic bulge implies that it is significantly more gamma-ray luminous than other gamma-ray detected novae ($L_{\gamma} \gtrsim 2 \times 10^{36}$ erg s$^{-1}$ in the energy range 100 MeV--10 GeV, exceeding other novae by $\gtrsim$ order of magnitude; \citealt{2015ApJ...809..160F, 2014Sci...345..554A}). We can also use this distance estimate to infer the nature of the nova host system; no underlying binary is detected in the VVV survey down to $m_K < 16.6$ mag \citep{Minniti10}, implying that the binary contains a dwarf star and not a red giant \citep{2015ApJ...809..160F}.

While previous work has established the context of V1324~Sco, our goal here is to present a detailed picture of the nova outburst itself, to understand how gamma-ray producing shocks formed. Section~\ref{sec:OpticalPhotometry} gives an overview of the outburst, as traced by UV/optical/near-IR (UVOIR) photometry.  In Section~\ref{sec:RadioSection} we present our radio data, and discuss a first peak in the radio light curve that is a strong indicator of shocks. We then use the second peak in the radio light curve to estimate the ejecta mass and kinetic energy of the outburst. In Section~\ref{sec:OpticalSpectra} we present the optical spectra and use them to constrain the kinematics and filling factor of V1324~Sco's outburst. In Section~\ref{sec:XRay} we detail the X-ray limits and discuss how they are consistent with observations at other wavelengths. In Section~\ref{sec:Discussion} we summarize that V1324 Sco is---in all non-gamma-ray observations---a classical nova. We discuss how classical novae can produce a range of gamma ray luminosities, and the conditions that might lead to the highest gamma-ray luminosity observed from a nova to date. Finally, we conclude the paper in Section~\ref{sec:Conclusion} by summarizing our findings for V1324~Sco.

\begin{deluxetable*}{cccccccc}
\tabletypesize{\scriptsize}
\tablecaption{\label{tab:photodata} Table of Photometric Data}
\tablehead{
\colhead{Observation Date}&
\colhead{JD}&
\colhead{$t-t_0$\tablenotemark{a}}&
\colhead{Filter}&
\colhead{Mag}&
\colhead{Mag Error}&
\colhead{Observer/Group}&
\colhead{Telescope/Specific Filter\tablenotemark{b}} \\
&& \colhead{(Days)}& &&&&
}
\startdata	
2012 Apr 13 & 2456030.07502 & -48.92499   & I & 18.700 & 0.150 & MOA & MJUO-Ibroad \\ 
2012 Apr 13 & 2456030.95591 & -48.04410   & I & 18.770 & 0.090 & MOA & MJUO-Ibroad \\ 
2012 Apr 13 & 2456030.95714 & -48.04287   & I & 18.590 & 0.090 & MOA & MJUO-Ibroad \\ 
2012 Apr 13 & 2456030.99663 & -48.00338   & I & 18.800 & 0.110 & MOA & MJUO-Ibroad \\ 
2012 Apr 14 & 2456031.05194 & -47.94807   & I & 18.760 & 0.090 & MOA & MJUO-Ibroad \\ 
2012 Apr 14 & 2456031.06304 & -47.93697   & I & 18.900 & 0.110 & MOA & MJUO-Ibroad \\ 
2012 Apr 14 & 2456031.07414 & -47.92587   & I & 18.850 & 0.090 & MOA & MJUO-Ibroad \\ 
2012 Apr 14 & 2456031.08652 & -47.91348   & I & 18.860 & 0.110 & MOA & MJUO-Ibroad \\ 
2012 Apr 14 & 2456031.09762 & -47.90238   & I & 19.030 & 0.110 & MOA & MJUO-Ibroad \\ 
2012 Apr 14 & 2456031.10976 & -47.89024   & I & 18.660 & 0.070 & MOA & MJUO-Ibroad \\ 
2012 Apr 14 & 2456031.12211 & -47.87789   & I & 18.930 & 0.120 & MOA & MJUO-Ibroad \\ 
2012 Apr 14 & 2456031.13321 & -47.86679   & I & 18.890 & 0.100 & MOA & MJUO-Ibroad \\ 
2012 Apr 14 & 2456031.14435 & -47.85566   & I & 18.830 & 0.100 & MOA & MJUO-Ibroad \\ 
2012 Apr 14 & 2456031.15670 & -47.84331   & I & 18.860 & 0.090 & MOA & MJUO-Ibroad \\ 
2012 Apr 14 & 2456031.16781 & -47.83220   & I & 18.950 & 0.110 & MOA & MJUO-Ibroad \\ 
2012 Apr 14 & 2456031.17893 & -47.82108   & I & 18.890 & 0.090 & MOA & MJUO-Ibroad \\ 
2012 Apr 14 & 2456031.19127 & -47.80874   & I & 18.910 & 0.090 & MOA & MJUO-Ibroad \\ 
2012 Apr 14 & 2456031.20237 & -47.79764   & I & 18.880 & 0.120 & MOA & MJUO-Ibroad \\ 
2012 Apr 14 & 2456031.21348 & -47.78653   & I & 18.850 & 0.090 & MOA & MJUO-Ibroad \\ 
2012 Apr 14 & 2456031.22585 & -47.77416   & I & 18.880 & 0.100 & MOA & MJUO-Ibroad \\ 
2012 Apr 14 & 2456031.23825 & -47.76176   & I & 18.750 & 0.090 & MOA & MJUO-Ibroad \\ 
2012 Apr 14 & 2456031.25182 & -47.74818   & I & 18.870 & 0.090 & MOA & MJUO-Ibroad \\ 
2012 Apr 14 & 2456031.96000 & -47.04001   & I & 18.610 & 0.080 & MOA & MJUO-Ibroad \\ 
2012 Apr 14 & 2456031.96123 & -47.03877   & I & 18.700 & 0.080 & MOA & MJUO-Ibroad \\ 
2012 Apr 14 & 2456031.99908 & -47.00093   & I & 18.880 & 0.120 & MOA & MJUO-Ibroad \\ 
2012 Apr 15 & 2456032.05184 & -46.94817   & I & 18.740 & 0.100 & MOA & MJUO-Ibroad \\ 
2012 Apr 15 & 2456032.07405 & -46.92596   & I & 18.840 & 0.130 & MOA & MJUO-Ibroad \\ 
2012 Apr 15 & 2456032.08742 & -46.91258   & I & 18.920 & 0.100 & MOA & MJUO-Ibroad \\ 
2012 Apr 15 & 2456032.09855 & -46.90146   & I & 18.810 & 0.100 & MOA & MJUO-Ibroad \\ 
2012 Apr 15 & 2456032.10966 & -46.89035   & I & 18.770 & 0.090 & MOA & MJUO-Ibroad \\ 
2012 Apr 15 & 2456032.12200 & -46.87801   & I & 18.880 & 0.150 & MOA & MJUO-Ibroad \\ 
2012 Apr 15 & 2456032.13311 & -46.86690   & I & 18.740 & 0.100 & MOA & MJUO-Ibroad \\ 
2012 Apr 15 & 2456032.14421 & -46.85580   & I & 18.850 & 0.100 & MOA & MJUO-Ibroad \\ 
2012 Apr 15 & 2456032.15655 & -46.84346   & I & 18.740 & 0.090 & MOA & MJUO-Ibroad \\ 
2012 Apr 15 & 2456032.16869 & -46.83132   & I & 18.780 & 0.090 & MOA & MJUO-Ibroad \\ 
2012 Apr 15 & 2456032.17989 & -46.82012   & I & 18.910 & 0.090 & MOA & MJUO-Ibroad \\ 
2012 Apr 15 & 2456032.19224 & -46.80777   & I & 18.960 & 0.100 & MOA & MJUO-Ibroad \\ 
2012 Apr 15 & 2456032.20335 & -46.79666   & I & 18.890 & 0.080 & MOA & MJUO-Ibroad \\ 
2012 Apr 15 & 2456032.21445 & -46.78556   & I & 18.800 & 0.080 & MOA & MJUO-Ibroad \\ 
2012 Apr 15 & 2456032.22683 & -46.77317   & I & 18.840 & 0.070 & MOA & MJUO-Ibroad \\ 
2012 Apr 15 & 2456032.23920 & -46.76081   & I & 18.840 & 0.100 & MOA & MJUO-Ibroad \\ 
2012 Apr 15 & 2456032.25030 & -46.74971   & I & 18.820 & 0.080 & MOA & MJUO-Ibroad \\ 
2012 Apr 15 & 2456032.26264 & -46.73736   & I & 18.960 & 0.190 & MOA & MJUO-Ibroad \\ 
... & ... & ...    & ... & ... & ... & ... & ... \\ 
\enddata
\tablecomments{All of these data, as well as data from AAVSO and~\citet{2012PASP..124.1057W}, can be found online.}
\tablenotetext{a}{Taking $t_0$ to be 2012 June 1.0}
\tablenotetext{b}{We abbreviate the different facilities used by the MOA and MicroFUN groups as: \textbf{MJUO}: Mt. John University Observatory; \textbf{AUCK}: Auckland Observatory; \textbf{CTIO}: SMARTS 1.3 Meter Telescope.}
\end{deluxetable*}

\section{UVOIR Photometry} \label{sec:OpticalPhotometry}
In this section, we present the UVOIR light curve and describe the basic phases of the nova outburst evolution for V1324~Sco.

\subsection{Observations and Reduction}
V1324 Sco falls within one of the fields that the MOA  Collaboration continually observes with the MOAII 1.8 meter telescope at Mt. Johns Observatory in New Zealand. V1324 Sco was initially detected in 2012 April by their high-cadence $I$-band photometry~\citep{2012ATel.4157....1W}. The initial detection showed a slow monotonic rise in brightness between April 13 - May 31, followed by a very large increase in brightness starting June 1 (Figure~\ref{fig:OpticalLC}; \citealt{2012ATel.4157....1W}). For the rest of this paper, we take 2012 June 1 to be day 0, or the start of the nova outburst. We also adopt the convention throughout this paper that all dates with $-$ or $+$ denote days before or after 2012 June 1, respectively.

All initial high-cadence observations, taken as part of the regular MOA program, were taken in the $I$-broad band, and were reduced using standard procedures (see~\citealt{2001MNRAS.327..868B} for details). The MOA survey emphasizes rapid imaging of the Galactic bulge fields; on a clear night an individual field will be imaged every $\sim$40 minutes. The result of this high time cadence photometry can be seen in Figure~\ref{fig:OpticalLC}. It should be noted that the primary purpose of the high-cadence observations is difference imaging; as a result, the individual values should only be used to measure changes, not as an absolute measurement~\citep{2001MNRAS.327..868B}. 

After the steep optical rise a follow-up campaign was triggered by the MicroFUN group\footnote{\url{http://www.astronomy.ohio-state.edu/~microfun/}}, who believed that the transient was a potential microlensing event. Apart from the standard $I$-broad band filter, the MicroFUN follow up observations also used $V$ and $I$ Bessel filters. Other observations were made in $B$, $V$, and $I$ filters using the Small \& Moderate Aperture Research Telescope System (SMARTS) 1.3 Meter telescope and Auckland Observatories. 

Along with the MOA and MicroFUN data we also present multi-color photometry from Fred Walter's ongoing Stony Brook/SMARTS Atlas of (mostly) Southern Novae (see~\citealt{2012PASP..124.1057W} for further information on this dataset), as well as data from American Association of Variable Star Observers (AAVSO)\footnote{\url{https://www.aavso.org/data-download}}. The SMARTS data uses the ANDICAM instrument on the 1.3 meter telescope, and provide both optical ($B$, $V$, $R$, $I$) and near-IR ($J$, $H$, $K$) filters going from day $+35$ to day $+124$, while the AAVSO data use optical ($V$, $B$, $R$) filters, and go from day $+7$ to day $+445$. 

Finally, we incorporate the UV data taken contemporaneously with the X-ray observations. The UV data comes from the Ultraviolet/Optical Telescope (UVOT; see \citealt{2005SSRv..120...95R} for further details) on board \emph{Swift}. Each observation was taken using the UVM2 filter, which is centered on 2246~\AA\ and has a FWHM of 498~\AA\ \citep{2008MNRAS.383..627P}. These observations were taken at the same time as the X-ray observations (see Section~\ref{sec:XRay}), stretching from day $+22$ to day $520$; however, we only include observations where V1324 Sco was detected.

\begin{figure*}[ht]
\includegraphics[width=\textwidth]{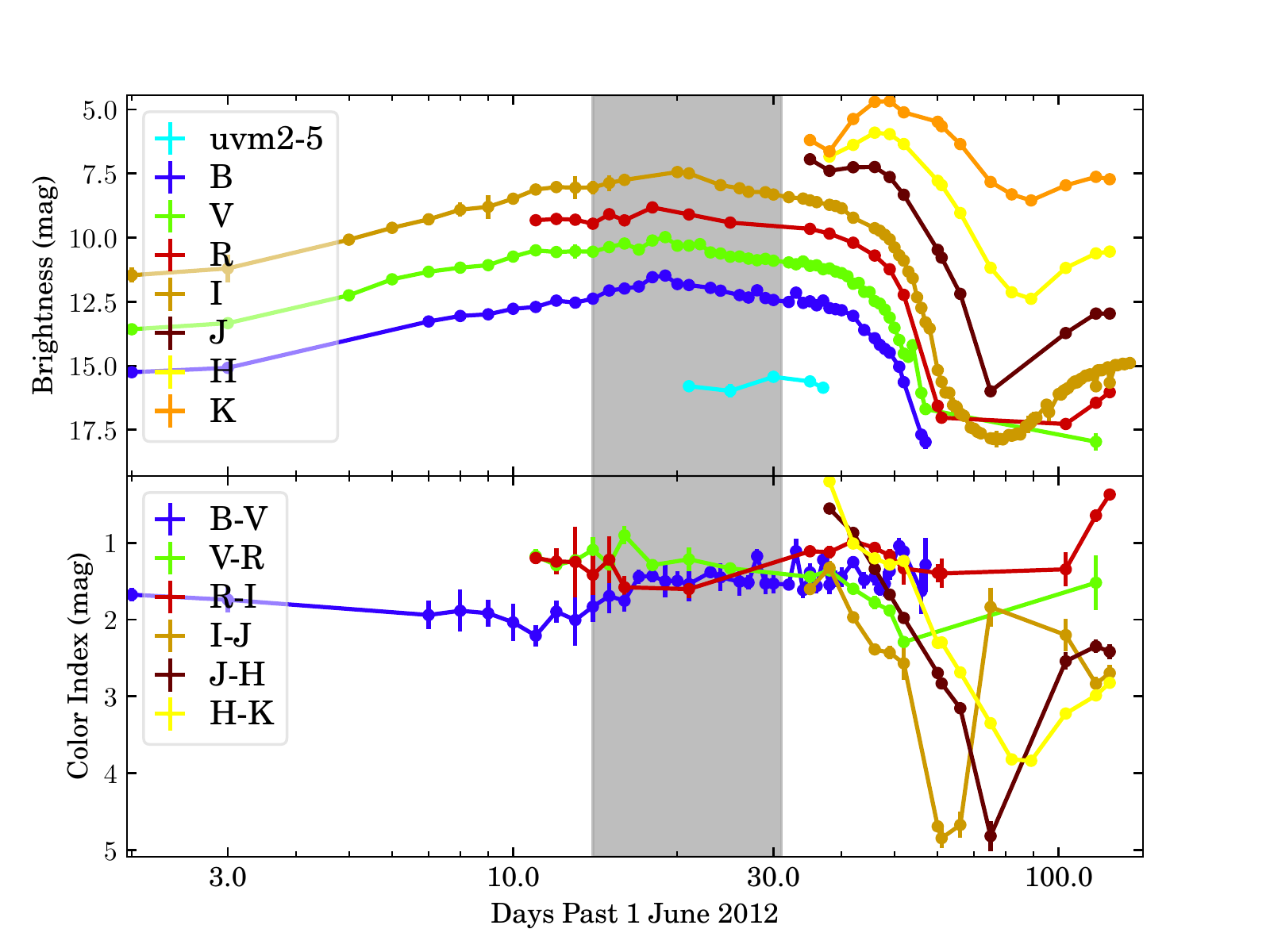}
\caption{\label{fig:OpticalColors}
Top panel: Light curves of V1324 Sco in the optical/near-IR bands. Bottom panel: Evolution of optical and near-IR colors. The gray shaded region denotes the time period when V1324 Sco was detected in gamma rays. Using this figure we can see how the dust event hits the bluer bands first and then moves to redder wavelengths as time progresses. We can also see that the dust event caused a drop in brightness all the way out to the near-IR ($JHK$) wavelength regime.
}\end{figure*}

A portion of the UVOIR data set is presented in Table~\ref{tab:photodata}; the entire data set can be found in the online publication. Multi-band photometry is plotted in Figure \ref{fig:OpticalColors}. Note that no attempt has been made to standardize the photometry from different observatories.

\subsection{Timeline of the Optical Light Curve}
\label{sec:OpticalTimeline}
We present an overview of the different phases in the evolution of the optical light curve, to help orient the reader to the different qualitative variations. These different phases come from the classification scheme laid out in~\cite{2010AJ....140...34S}---with the exception of the early time rise. Throughout this overview we will reference Figure~\ref{fig:OpticalLC} and Figure \ref{fig:OpticalColors};
note that while Figure \ref{fig:OpticalColors} features multiple bands, it has significantly lower time resolution than Figure~\ref{fig:OpticalLC}.

\subsubsection{Early-Time Rise (Days $-$49 to 0)}
The first MOA $5\sigma$ detection of V1324 Sco occurred on 2012 April 13. Following this, there was a monotonic increase in brightness that lasted until 31 May 2012. The total increase in brightness during this period was $\Delta I \approx 2.5$ mags (about $\sim 0.05$ mags per day). This early-time rise can be seen as phase A of Figure~\ref{fig:OpticalLC}. 

This type of early-time rise has been observed twice before---in V533 Her and V1500~Cyg~\citep{1975AJ.....80..515R, Collazzi09}---but no theory has been put forward to explain the phenomenon. It is worth noting that most novae lack pre-eruption photometry. Sky monitors like the Solar Mass Ejection Imager (SMEI)~\citep{2010ApJ...724..480H} and ASAS-SN \citep{Shappee14} have relatively shallow limiting magnitudes, preventing them from seeing such faint early time rises. It is only with the type of dedicated, deep, high cadence observations like those of MOA that we can observe such a rise.
Catching such an early time rise is unusual, and deserves a thorough analysis that goes beyond the scope of this paper. We therefore defer the discussion of this period to Wagner et al. (2017, in preparation).

\subsubsection{Onset of the Steep Optical Rise (Days 0 to $+$10)} 
The slow monotonic rise was followed by a rapid increase in brightness; between day 0 and day $+2$ the brightness increased by $\sim 2.2~\mathrm{mag}~\mathrm{day}^{-1}$. 
Between days $+2.8$ and $+3.3$ the rate of increase dropped to $\sim 1.1~\mathrm{mag}~\mathrm{day}^{-1}$, and then between days $+5.6$ and $+9.2$ the rate dropped further to $\sim 0.3~\mathrm{mag}~\mathrm{day}^{-1}$. 

The next time V1324~Sco was visited, on day $+$12.9, the light curve appears to have flattened out. During the period, day 0 to day +9.2, the I band flux increased by a total of $\sim$9.1 magnitudes, with most of that rise occurring during the first $\sim$3 days. This rise can be seen as phase B of Figure \ref{fig:OpticalLC}. Note that the large uncertainty in measurement on day $+$2 is the result of binning the measurements during the steep optical rise.

\subsubsection{Flattening of the Optical Light Curve (Days $+$10 to $+$45)} 
The dramatic increase in the optical flux was followed by a period with a much smaller change in brightness. This flattening in the light curve is not unique to V1324~Sco; \citet{2010AJ....140...34S} show 15 examples of nova light curves with a similar flattening around peak, 10 of which also show a dust event. This ``flat top" can be seen as phase C of Figure \ref{fig:OpticalLC}.

Note that the apparent fluctuations in the MOA light curve during this period are likely an artifact of observing an unusually bright source (i.e., saturation). The light curve around maximum is better represented by the CTIO photometry shown in Figure \ref{fig:OpticalColors}. The light curve shows a very gradual, gentle rise to a maximum, $I = 8.2$ mag on day $+$21. 
The light curve then gradually decreases until about day $+$45, when a rapid decrease in flux is brought on by a dust event.

It is during the flattening of the optical light curve that we see both the gamma-ray emission as well as the beginning of the initial radio bump (see section~\ref{sec:DiscussionGammaRayNova} for further details).

\subsubsection{Dust Event (Days $+$45  to $+$157)} 
The flattening of the optical light curve was followed by another rapid change in brightness, this time downwards. There was a very clear steep decline in optical and near-IR flux that took place from day $+46$ to day $+78$, and a subsequent recovery from day $+79$ to day $+157$. Only the MOA $I$ band data had the cadence and sensitivity necessary to capture the minimum of the decrease; the $I$-band flux dropped by $\sim 8.5$ magnitudes in the span of $\sim$ 30 days (Phase D in Figure~\ref{fig:OpticalLC}). 
Figure~\ref{fig:OpticalColors} shows that this decline in flux occurred all the way out to the near-IR (although the decrease was much less in the near-IR bands, i.e. only $\sim 3.9$ mags in $K$ band). This decline in flux that preferentially affects the bluer light is the signature of a nova dust event.

A dust event occurs in a nova when the ejecta achieve conditions that are conducive to the condensation of dust---e.g., cool, dense, and shielded from ionizing radiation \citep{1977AJ.....82..209G, 1988ARA&A..26..377G}. The newly formed dust has a large optical depth; as a result a new, cooler, photosphere is created at the site of dust condensation. 


\subsubsection{Power Law Decline (Days $+157$ to End of Monitoring)} Following the post-dust event rebound, the magnitude evolution followed a power law decline, with $I \propto (t-t_0)^{0.2}$ (where $t_0$ is 2012 June 1).
This decline continues until the final observation from April 2014, when V1324~Sco fell below the MOA detection threshold. In Figure~\ref{fig:OpticalLC}, the power-law decline is phase E, between day $+228$ and $+730$.

\subsection{Discussion of the UVOIR light curve}

In the optical regime, V1324 Sco is photometrically a \emph{D} (Dusty) class nova \citep{2010AJ....140...34S}, because of the extraordinary dust event that took place between days $+46$ to $+157$. Other \emph{D} class novae include FH Ser, NQ Vul, and QV Vul \citep{2010AJ....140...34S}. Among $D$-class novae, the speed of V1324~Sco's photometric decline is quite typical. V1324~Sco's $t_2$ value---that is, the time for a nova to decline by 2 magnitudes from maximum in $V$ band---is $t_2 \approx 24$ days. This is consistent with other \emph{D} class novae, all of which are of order tens of days 
(see \citealt{2010AJ....140...34S} and references therein).

In the case of V1324~Sco, the dust event includes a drop in flux all the way out to the near-IR. This suggests that the change in temperature from optical maximum was significant, and that the dust photosphere was cold. A fit to the near-IR colors at the epoch closest to the $I$ band minimum suggest that the dust photosphere was $<1000$ K.  While dust events are quite common in novae---\cite{2010AJ....140...34S} gives 16 examples of other such novae---there are only a few novae with dust dips showing comparably cool photospheres (e.g. QV Vul and V1280 Sco;~\citealt{1992ApJ...400..671G, 2015arXiv150708801S}).

If the shocks in novae are dense and radiative (as predicted by \citealt{2014MNRAS.442..713M, 2015MNRAS.450.2739M}), then they are ideal locations for dust formation \citep{2016arXiv161002401D}. Radiative shocks can also explain the observed gamma-ray luminosity and non-detection in X-rays (Section \ref{sec:XRay}; \citealt{2014MNRAS.442..713M}). When compared to other gamma-ray detected novae in the AAVSO database \citep{Kafka16}, V1324~Sco had an unusually dramatic dust event. For example, there is no sign of dust formation in V959~Mon \citep{Munari15}, and V339~Del showed signatures of dust formation at infrared wavelengths, but the event did not have a profound effect on the optical light curve \citep{Gehrz15, Evans17}. \citet{2016arXiv161002401D} find that some of these variations could be attributable to viewing angle, if dust preferentially forms along the orbital plane as would be expected in the geometry suggested by \cite{2014Natur.514..339C}. In addition, \cite{Evans08} point out that novae on CO white dwarfs are more likely to produce dust than novae on ONe white dwarfs. Observations of dust (or lack thereof) in these three gamma-ray detected novae can be reconciled if V1324~Sco is viewed at an edge-on inclination and hosts a CO white dwarf, while V959~Mon's binary hosts a ONe white dwarf viewed at high inclination \citep{Page13_v959mon, Shore13}, and V339~Del hosts a CO white dwarf but is observed at low inclination \citep{Schaefer14, Shore16}.

The origin of ``flat tops'' in novae remains something of an open question. In luminous red novae like V1309~Sco (which eject several orders of magnitude more mass than classical novae), \citet{2013Sci...339..433I} proposed that a plateau around maximum is caused by a recombination front, much as in Type IIP supernovae \citep{Chugai91}. In this case, the light curve flattening near maximum is explained as a  photosphere radius that does not change substantially in Eulerian coordinates (but shrinks in Lagrangian coordinates) and has roughly constant temperature, due to the fact that the ejecta are cooling and recombining. 

Another possible explanation for flat-topped light curves was developed in the case of T~Pyx and proposed by \citet{2014ApJ...785...78N} and \citet{Chomiuk14}, where there is multi-wavelength evidence that the bulk of the ejecta remained in a quasi-hydrostatic configuration around the binary until the end of the ``flat top'' period. In this nova, it appears that 1--2 months pass before the ejecta are accelerated to their terminal velocity and are expelled from the environs of the binary, although the physical origin of the delay remains a mystery (it is, perhaps, attributable to binary interaction with the quasi-static envelope). 

It should be noted that, of the gamma-ray detected novae, at least two---V1369 Cen and V5668 Sgr---had similar flattening of the optical light curve near maximum~\citep{2016ApJ...826..142C}, though both exhibited large ($\Delta V > 1$ Mag) fluctuations in brightness during their period of flattening (unlike V1324~Sco; Figure \ref{fig:OpticalColors}). In both systems, the evolution of optical spectral line profiles around maximum imply that several episodes of mass ejection transpire over the course of the variegated plateau \citep{2012PASP..124.1057W}. These systems support the idea that flat-topped light curves in novae may be a signature of complex, prolonged mass loss---the sort of mass loss which will produce shocks and gamma rays.

\section{Radio Data}
\label{sec:RadioSection}
Radio emission from novae is a crucial tool in understanding nova energetics, as the opacity at radio frequencies is directly proportional to the emission measure of the ionized ejecta---defined for some line of sight $z$ as $EM_z =\int n_e^2 dz$. Therefore, we can map out the density profile of the ejecta just by watching the evolution of the radio emission~\citep{1977ApJ...217..781S, 1979AJ.....84.1619H, Seaquist_Bode08, Roy12}. The early time radio light curve can also show unexpected behavior that can be used to constrain shocks in the nova event \citep{1987A&A...183...38T, Krauss11, 2014Natur.514..339C, 2016MNRAS.457..887W, 2016MNRAS.460.2687W}. 

\begin{figure*}[ht]
\includegraphics[width=2.0\columnwidth]{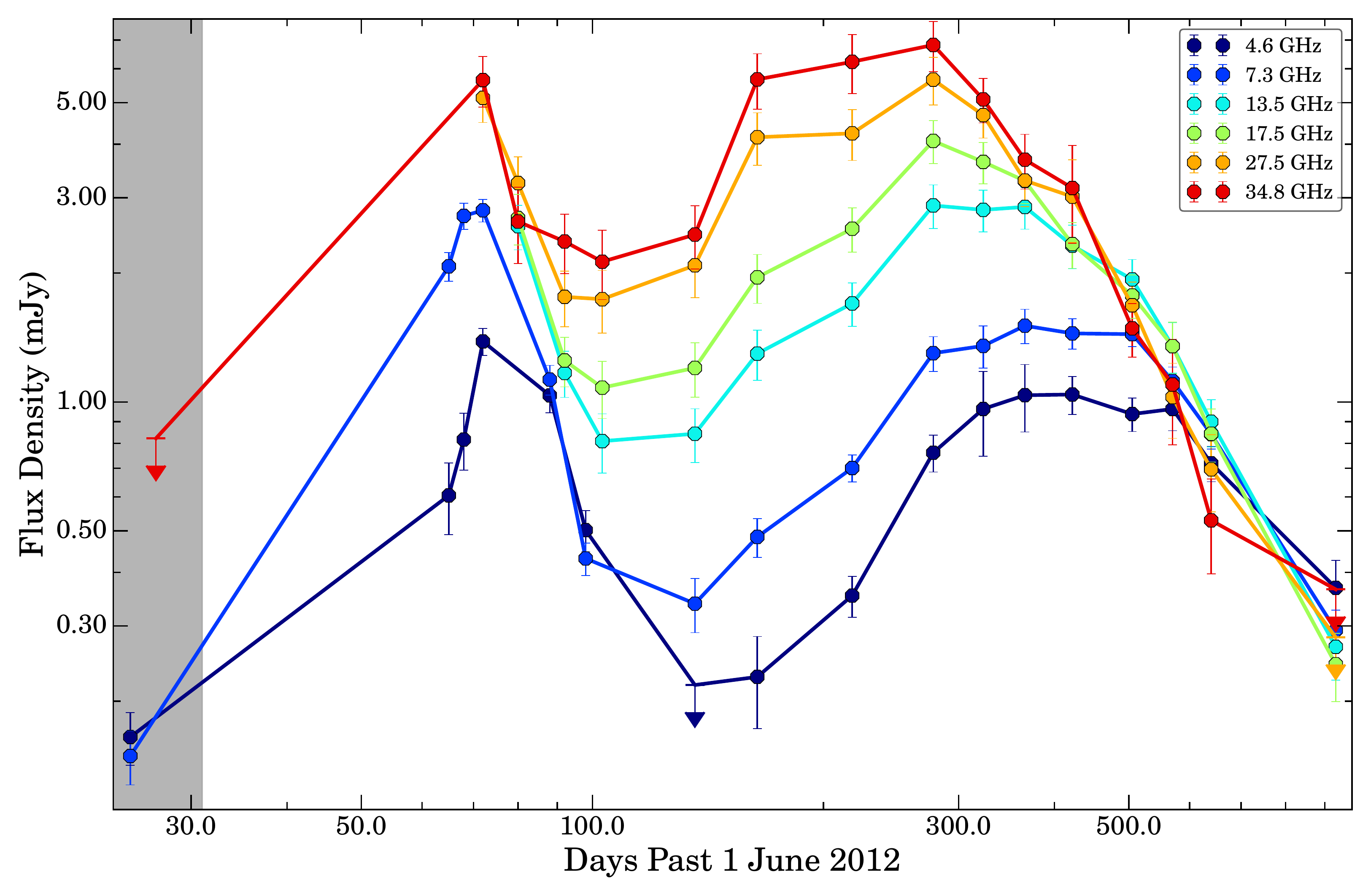}
\caption{\label{fig:RadioLC}
Radio light curve for V1324 Sco, spanning day $+22$ to day $+930$ (using June 1 2012 as day 0). The initial maximum takes place between day $+25$ to day $+136$, while the second radio peak occurs around day $+300-400$. The time range of GeV gamma-ray detections is highlighted in grey. 
}\end{figure*}

\begin{figure*}[ht]
\includegraphics[width=\textwidth]{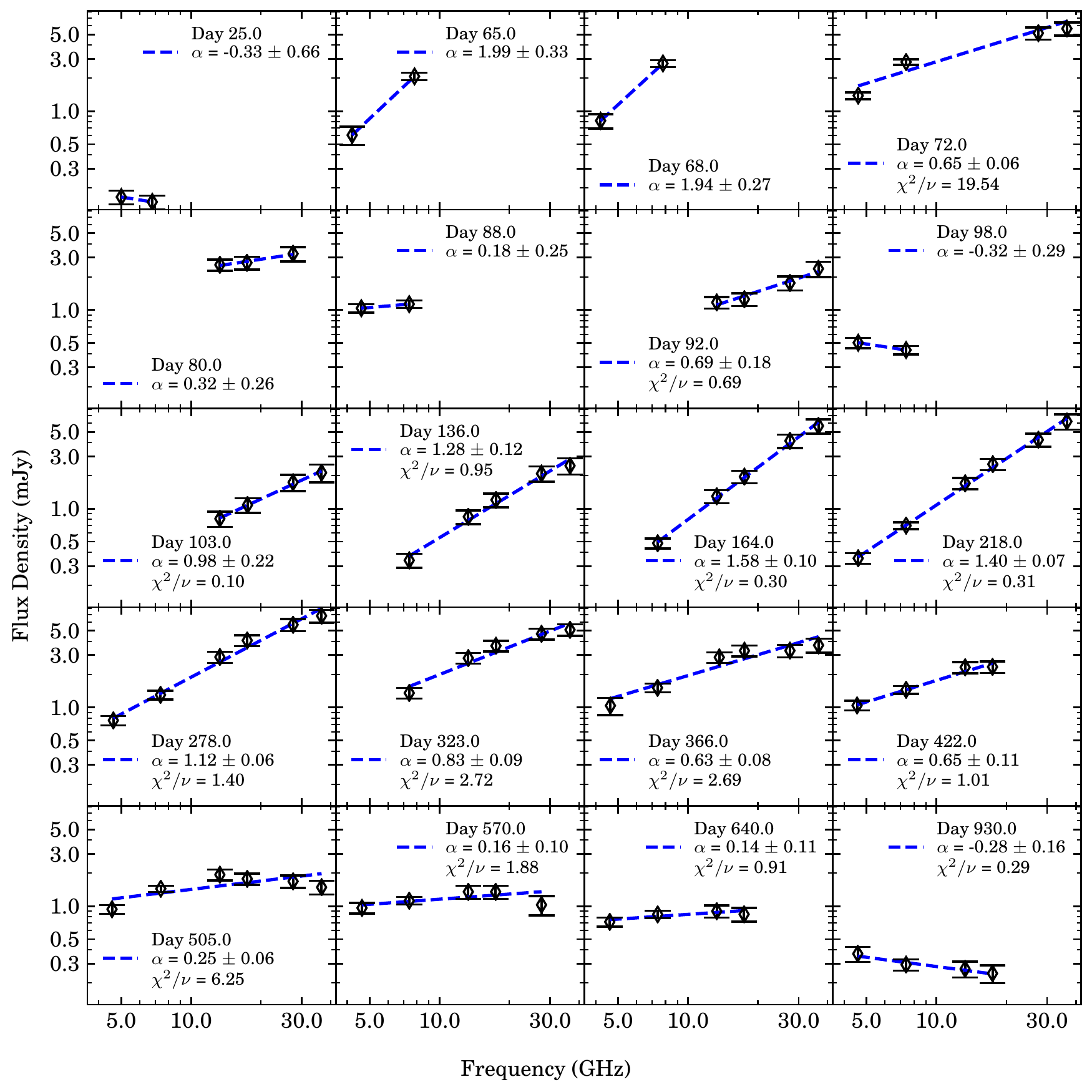}
\caption{\label{fig:RadioSpectra}
The evolution of the radio spectral energy distribution for V1324 Sco. At every epoch with measurements at three or more frequencies we fit either a power-law or double power-law to the flux values. The best fit solution was selected based on reduced chi-squared value closest to 1.  
}\end{figure*}

\subsection{Observations and Reduction}
We obtained sensitive radio observations of V1324~Sco between 2012 June 26 and 2014 December 19 with the Karl G. Jansky Very Large Array (VLA) through programs S4322, 12A-483, 12B-375, 13A-461, 13B-057, and S61420.  Over the course of the nova, the VLA was operated in all configurations, and data were obtained in the C (4--8 GHz), Ku (12--18 GHz), and Ka (26.5--40 GHz) bands, resulting in coverage from 4--37 GHz.  Observations were acquired with 2 GHz of bandwidth and 8-bit samplers, split between two independently tunable 1-GHz-wide basebands. The details of our observations are given in Table~\ref{tab:RadioObs}. 

At the lower frequencies (C band), the source J1751-2524 was used as the complex gain calibrator, while J1744-3116 was used for gain calibration at the higher frequencies (Ku and Ka bands). The absolute flux density scale and bandpass were calibrated during each run with either 3C48 or 3C286. Referenced pointing scans were used at Ku and Ka bands to ensure accurate pointing; pointing solutions were obtained on both the flux calibrator and gain calibrator, and the pointing solution from the gain calibrator was subsequently applied to our observations of V1324 Sco. Fast switching was used for high-frequency calibration, with a cycle time of $\sim$2 minutes.  Data reduction was carried out using standard routines in AIPS and CASA \citep{Greisen03, McMullin07}. Each receiver band was edited and calibrated independently. The calibrated data were split into their two basebands and imaged, thereby providing two frequency points. 

An observation in A configuration (the most extended VLA configuration) from 2012 Dec 16 suffered severe phase decorrelation at higher frequencies. Despite efforts to self calibrate, we could not reliably recover the source and we therefore do not include these measurements here. 

In each image, the flux density of V1324 Sco was measured by fitting a Gaussian to the imaged source with the tasks \verb|JMFIT| in AIPS and \verb|gaussfit| in CASA. We record the integrated flux density of the Gaussian; in most cases, there was sufficient signal on V1324 Sco to allow the width of the Gaussian to vary slightly, but in cases of low signal-to-noise ratio, the width of the Gaussian was kept fixed at the dimensions of the synthesized beam. Errors were estimated by the Gaussian fitter, and added in quadrature with estimated calibration errors of 5\% at lower frequencies ($<$10 GHz) and 10\% at higher frequencies ($>$10 GHz).  All resulting flux densities and uncertainties are presented in Table \ref{tab:RadioObs}. V1324 Sco appeared as an unresolved point source in all observations.

Next, we discuss the different phases of the radio light curve evolution. The radio emission is shown in Figure~\ref{fig:RadioLC} (radio light curve) and Figure~\ref{fig:RadioSpectra} (radio spectral energy distributions). 

\subsection{Initial Radio Maximum and Shock Emission}
\label{sec:FirstRadioBump}
V1324 Sco was detected during the first radio observation (day $+25$), coincident with the end of the gamma-ray emission. In subsequent radio observations the light curve rose steeply to a first maximum, peaking on day $+72$. This initial peak is in contrast with the second maximum, which peaked around day $+300$ (see Figure~\ref{fig:RadioLC}).  

The radio spectrum on the rise to initial maximum started out consistent with flat (albeit with a large error bar): $\alpha = -0.3 \pm 0.7$ on day $+25$ (where $\alpha$ is defined as $f_{\nu} \propto \nu^{\alpha}$, $f_{\nu}$ is the flux density, and $\nu$ is the observing frequency; see Figure~\ref{fig:RadioSpectra}). The radio spectrum then rapidly transitioned to $\alpha = 2.0 \pm 0.3$ on day $+65$ (the time of initial maximum), implying optically thick emission. The spectrum then flattened out again ($\alpha = 0.6 \pm 0.1$ on day $+72$). 

During this initial radio peak, the light curve rises steeply, as $f_{\nu} \propto t^{2.9}$, assuming day 0 is 2012 June 1. This is steeper than expected for expansion of an optically-thick isothermal sphere ($f_{\nu} \propto t^{2}$; \citealt{Seaquist_Bode08}), and could indicate that this maximum is dominated by thermal emission increasing in temperature or non-thermal emission.

To further investigate the nature of this initial radio maximum, we can use the brightness temperature, which is a proxy for surface brightness. Brightness temperature parameterizes the temperature that would be necessary if the observed flux originated from an optically-thick thermal blackbody. The equation for brightness temperature is given by 
\begin{equation}
T_b(\nu, t) = \frac{S_\nu (t) c^2 D^2}{2 \pi k_b \nu^2 (v_{\mathrm{ej}}t)^2},
\end{equation}
where $S_\nu$ is the observed flux, $D$ is the distance, $t$ is the time since explosion, $v_{\mathrm{ej}}$ is the ejecta velocity, $\nu$ is the observing frequency, and $k_b$ is the Boltzmann constant. Typical brightness temperatures of thermal emission from novae are $\sim 10^4$ K \citep{2015ApJ...803...76C}. If the measured brightness temperature is far in excess of $\sim 10^4$ K, it is a solid indication of synchrotron emission.

\begin{figure}[h]
\includegraphics[width=\columnwidth]{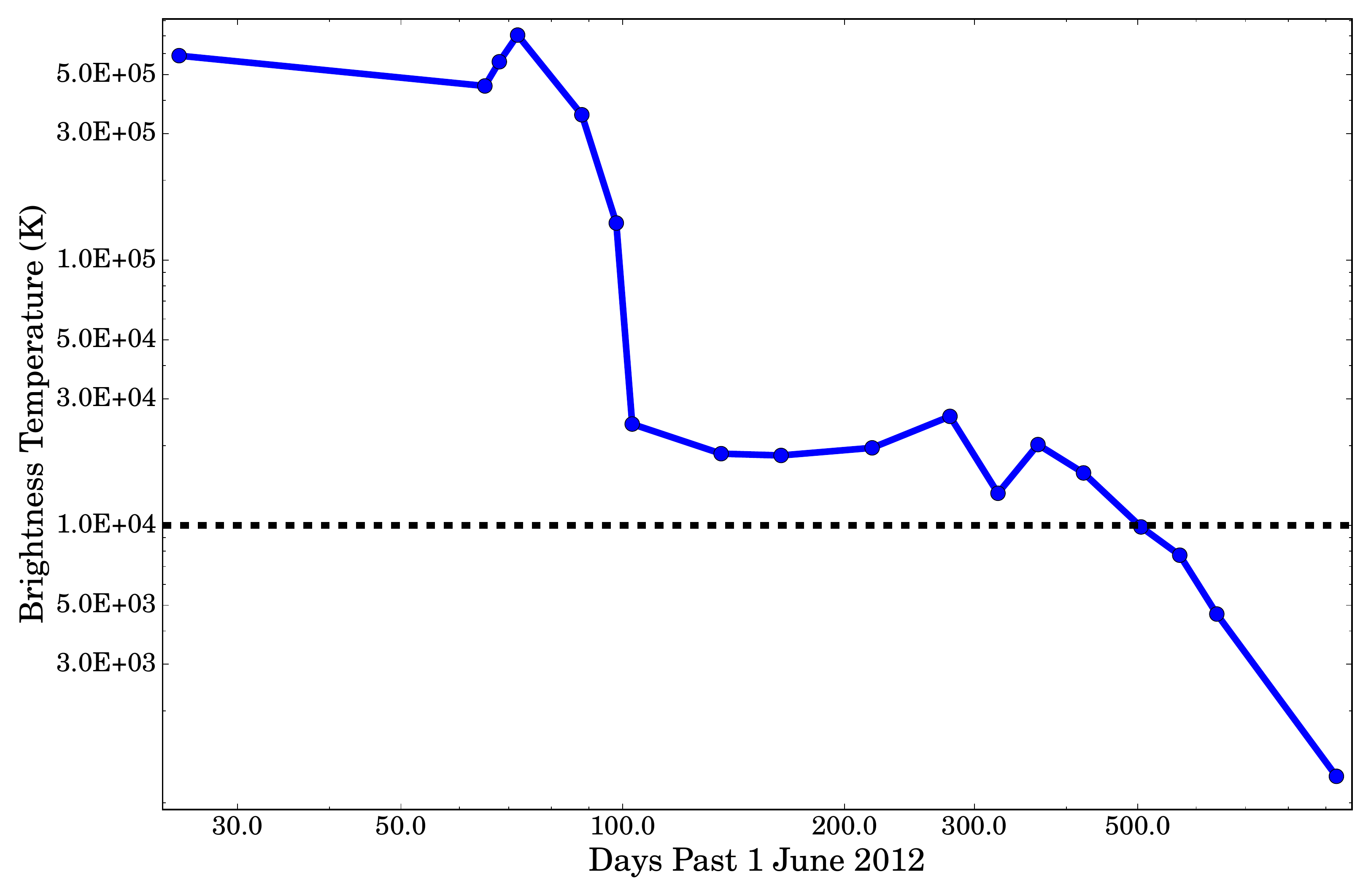}
\caption{\label{fig:BrightnessTemperature}
The early time radio bump (day $30$ to $50$) has maximum brightness temperatures far exceeding that of the canonical $10^4$ K thermally emitting ejecta---which can be seen as the dashed black line. The first radio observation, on day $+25$, occurred while the source was still gamma-ray bright. Note that the observation epochs taken on days $88$ and $92$ were omitted due to a lack of low-frequency observations.
}\end{figure}

Figure~\ref{fig:BrightnessTemperature} shows estimates of the brightness temperature as a function of time, using the distance lower limit of $6.5$ kpc from~\citealt{2015ApJ...809..160F} and a velocity of 1,000 km s$^{-1}$ (the velocity of the slow flow as estimated from optical spectroscopy; Section \ref{sec:VelocityVariations}). Two observation epochs (days $80$ and $92$) were removed due to the lack of low-frequency observations, which usually set the maximum brightness temperature. Note that a larger distance would increase these values, while a larger $v_{\mathrm{ej}}$ would decrease them. If we instead assume a velocity of 2,600 km s$^{-1}$ (the velocity of the fast flow; Section \ref{sec:VelocityVariations}), the brightness temperature estimates will decrease by a factor, $\sim$7.

During the initial radio maximum, we see that the brightness temperature substantially surpasses $10^4$ K, registering at $5 \times 10^5$ K (assuming 1,000 km s$^{-1}$ outflow velocity, and still $\sim 10^5$ K if we assume the faster outflow). Such brightness temperatures are very difficult to produce with thermal emission alone, and can be most easily explained as synchrotron emission \citep{1987A&A...183...38T, 2016MNRAS.460.2687W}. 

This type of initial radio bump has been seen in several other nova, including QU Vul~\citep{1987A&A...183...38T}, V1723 Aql~\citep{Krauss11, 2016MNRAS.457..887W}, and V5589 Sgr~\citep{2016MNRAS.460.2687W} and we are beginning to develop theories to explain such behavior. These previous works, along with theoretical analysis by \cite{2014MNRAS.442..713M} and \cite{2016MNRAS.463..394V}, postulate that the initial radio maxima could be either synchrotron emission or (unusually hot) thermal free-free emission. 

To explain the initial maximum as thermal emission, rather extraordinary conditions are needed. Both \cite{1987A&A...183...38T} and~\cite{2014MNRAS.442..713M} invoke a strong shock as a means for generating hot, free-free emitting gas. Note that the gas would not only need to be hot ($>10^5$ K), but also dense, as it would need to be optically thick to produce the initial maximum. Such large amounts of high temperature gas would also produce significant X-ray emission, which was not observed in V1324 Sco. This could be explained if there is a high column density of material absorbing the X-ray emitting region (see Section~\ref{sec:XRay}). However, for low-velocity ($v_{sh} \lesssim 1500 \kms$) shocks expanding into dense media, like internal shocks in novae, cooling is very efficient and drives the post-shock gas temperature to $T\sim 10^4$ K \citep{2014MNRAS.442..713M, 2016arXiv161002401D}. This makes it difficult to achieve the $\sim 10^5-10^6$ K gas necessary for the initial radio bump to be explained by thermal emission. 

A non-thermal explanation for the initial radio maximum is preferred by \cite{2016MNRAS.463..394V}, as synchrotron emission is an elegant explanation for brightness temperatures substantially in excess of $10^4$ K. A peak in the radio light curve could be produced by synchrotron emission suffering free-free absorption (or perhaps via the Razin-Tsytovich effect; see also \citealt{1987A&A...183...38T}). In this scenario, on the rise to maximum, the spectral index is predicted to be $\alpha = 2$, the light curve peaks when optical depth is of order unity, and during the optically-thin decline from maximum, the spectral index would be $\alpha = -0.5$ to $-1.0$ \citep{2016MNRAS.463..394V}. While such evolution of the spectral index is widely seen in supernovae \citep[e.g.,][]{Chevalier82, Weiler02}, the spectral index evolution during V1324~Sco's first radio maximum looks quite different. The spectral index never drops below $\alpha \approx 0.2$ (day $+$88; Figure \ref{fig:RadioSpectra}). See the top panel of Figure 14 in \citet{2016MNRAS.463..394V} for an illustration of how free-free absorbed synchrotron emission provides a problematic fit to the radio spectral energy distribution during the decline from V1324~Sco's initial radio maximum.

Similar radio spectral index evolution, combined with high brightness temperatures, have now been seen in several other novae, and a synchrotron explanation is favored over a thermal one \citep{1987A&A...183...38T, Krauss11, 2016MNRAS.460.2687W, 2016MNRAS.463..394V}. However, the physical explanation for a relatively flat (non-negative) spectral index on the decline from initial maximum, when the emission is expected to be optically thin, remains a mystery. Perhaps yet-unexplored physics is affecting the energy spectrum of non-thermal leptons in novae, making the spectrum flatter than predicted by models of diffusive shock acceleration \citep{Bell78, 1978ApJ...221L..29B}. Regardless of a thermal or synchrotron origin, the initial radio maximum in V1324~Sco is a clear indication of shocks in the months following outburst.

\subsection{Second Radio Maximum and Determination of Ejecta Mass}
\label{sec:RadioFitting}

After this initial radio bump, a second radio maximum occurred, starting sometime around September 15 2012 (day $+106$). It first appeared at high frequencies and progressed toward lower frequencies. During this second radio maximum, V1324 Sco peaked at $6.8$ mJy at high frequency (36 GHz) on day $+278$, and peaked at $\sim 1.0$ mJy for low frequency (4.5 GHz) on day $+422$.

The evolution of the second radio maximum is consistent with the ``standard'' picture of radio emission in novae---namely thermal emission from the $10^4$ K expanding ejecta \citep{1977ApJ...217..781S, 1979AJ.....84.1619H}. This portion of V1324~Sco's radio light curve is similar to the other novae that have been studied in the radio (e.g., \citealt{1977ApJ...217..781S,1979AJ.....84.1619H,2012ApJ...761..173C, 2014ApJ...785...78N,2016MNRAS.457..887W}).

The spectral index is steep on the rise to second maximum, reaching $\alpha = 1.6$ on day $+$164 (once there has been time for the initial radio bump to fade away). By day $+$323, there is clear evidence that the radio spectrum is flattening at higher frequencies (Figure \ref{fig:RadioSpectra}). This spectral turnover cascades to lower frequencies, until by day $+$640, the radio spectrum is consistent with optically-thin free free emission ($\alpha = -0.1$). This spectral index evolution is consistent with expectations for free-free emission from expanding thermal ejecta \citep{Seaquist_Bode08}.

The power-law rise to second maximum is also consistent with expectations for an isothermal sphere expanding at constant velocity ($f_{\nu} \propto t^2$; \citealt{Seaquist_Bode08}). The rise to second maximum at 7.5 GHz is well approximated by a power law with index 2 (assuming that day 0 is 2012 June 1). The rise to second maximum at 17.5 GHz is a bit shallower (power law index of 1.7), and this difference is likely attributable to the more substantial effect of the first radio maximum on the light curve between days $+$100--200 (Figure \ref{fig:RadioLC}). 

We therefore modeled the second radio bump as thermal emission from the expanding nova ejecta. We fit the radio data observed after day $+106$ using the standard model of \citet{1979AJ.....84.1619H}. Specifically, we utilize a homologously expanding ``Hubble flow" model, where the fastest ejecta are found at largest radii and throughout the ejecta, $v \propto r$. The ejecta are bounded at an inner and outer radius, and  we refer to the ratio between these as $\xi$. In between these inner and outer radii, we estimate an $r^{-2}$ density profile (for more details on this model, see \citealt{1977ApJ...217..781S}). The other physical quantities that go into the Hubble flow model are ejecta mass, maximum ejecta velocity, filling factor, temperature, and distance. More details on this radio light curve model and interplay between these variables can be found in Appendix A. 

\begin{figure*}[ht]
\includegraphics[width=2.0\columnwidth]{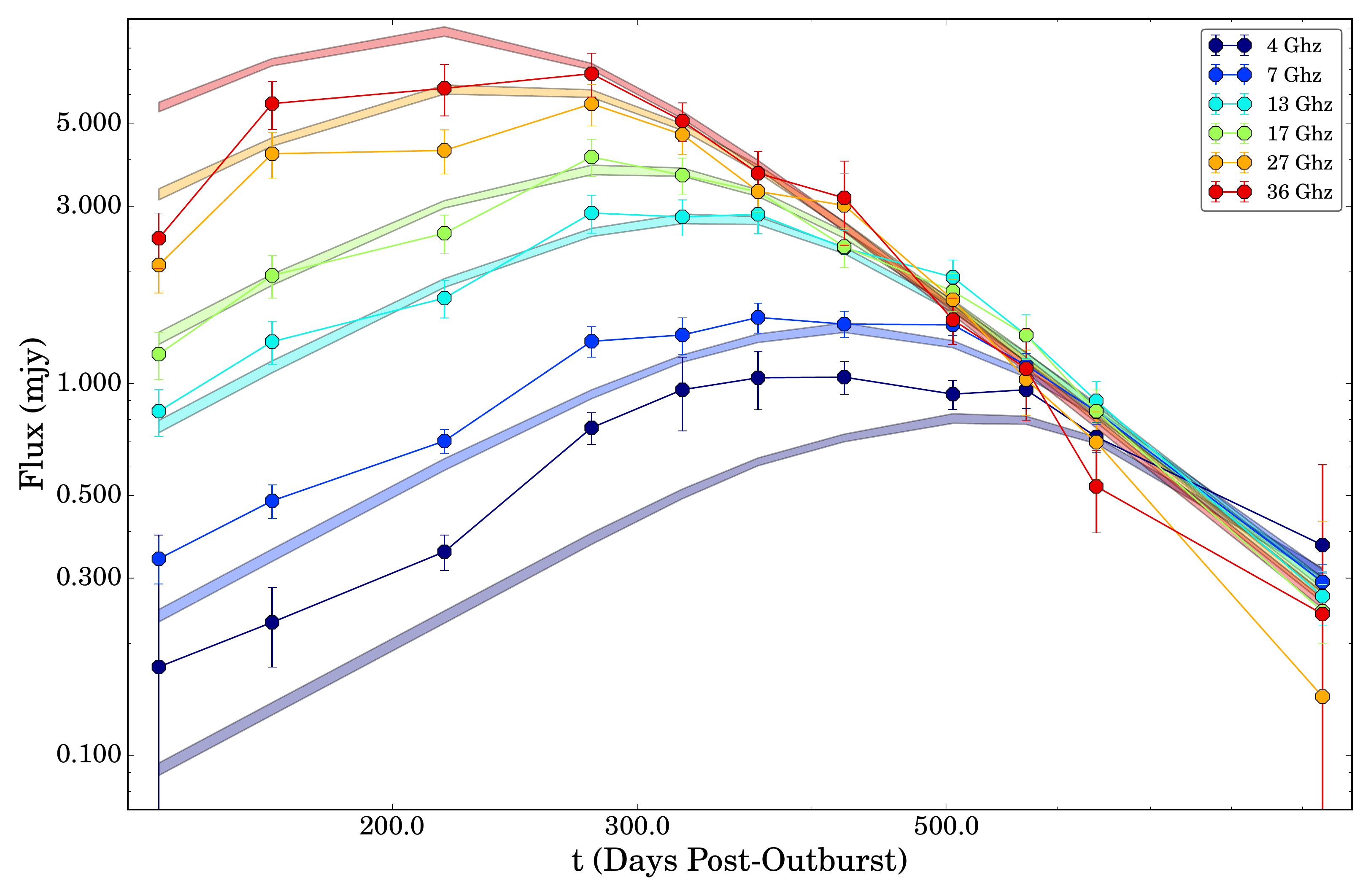}
\caption{\label{fig:RadioLC_WithFit}
Best fit model to \emph{just} the second bump portion of the radio light curve (see Figure~\ref{fig:RadioLC} for the entirety of the radio light curve). The best fit parameter values  and the resultant physical values can be found in Section~\ref{sec:RadioFitting}. The reduced chi-squared value fit for this model is $\chi^2_{red} = 3.36$. The fitting scheme was error weighted, which (partially) explains the relatively poor fits to the highest and lowest frequencies. There is the further issue of the fact that none of the data have the canonical optically thick spectral index of $\alpha = 2.0$, which the model expects (see Figure~\ref{fig:RadioSpectra}). 
}\end{figure*}

Figure~\ref{fig:RadioLC_WithFit} shows the fit to the second peak in the radio light curve using this model. The reduced chi-squared value fit for this model is $\chi^2/\nu = 3.36$. The fitting scheme was error weighted, which partially explains why the highest and lowest frequencies are not fit as well.  Further, by construction the model has a spectral index of $\alpha = 2.0$ during the rise, as this is the spectral index of optically thick thermal emission in the Rayleigh-Jeans tail. As can be seen in Figure~\ref{fig:RadioSpectra}, we never observe a spectral index this high; during the rise to second optical maximum, we observe $\alpha = 1.0-1.6$. This discrepancy between observed and predicted spectral index is common among novae \citep{Roy12, 2014Natur.514..339C, 2014ApJ...785...78N, Wendeln17}, and currently lacks a suitable explanation. Clearly, V1324~Sco is another data point illustrating that this discrepancy requires more attention.

Despite discrepancies in the spectral index on the rise, the flux density and timescale of the second peak reveal important information on the mass and energetics of the explosion. We can derive physical parameters for the ejecta---ejected mass ($M_{\rm ej}$) and total ejecta kinetic energy (${\rm KE}_{\rm ej}$), as well as the distance---from the model fit, with some assumptions. We assume the canonical temperature of photoionized gas---$10^4$ K~\citep{1989agna.book.....O,2015ApJ...803...76C}. This ejecta temperature is not only theoretically predicted, but has been observed in resolved radio images of novae \citep[e.g.,][]{1988Natur.335..235T, Hjellming96}. We also take the maximum ejecta velocity to be \AssumedEjectaVelocity (see Section~\ref{sec:SpectraEvol}), and a volume filling factor  of $f_V =$ \FillingFactor \  (see Appendix B).

The physical quantities derived from the light curve fit are:
\begin{eqnarray}
D = \DerivedDistance;\\ 
M_{\rm ej} = \EjectaMass;\\
{\rm KE}_{\rm ej} = \EjectaEnergy.
\end{eqnarray}
where the uncertainties quoted are 1$\sigma$ values. Uncertainty in filling factor have been propagated through this estimate and are included in the error bars. For both the derived distance and the total kinetic energy, the dominant source of uncertainty comes from $v_{\rm max}$; ${\rm KE}_{\rm ej}$ has a very strong dependence on the maximum velocity (${\rm KE}_{\rm ej} \propto v_{\rm max}^4)$. 
The uncertainty in the derived mass is dominated by the uncertainty in the filling factor measurement (although uncertainty in $v_{\rm max}$ is still non-negligible).

Let us now consider how these derived values depend on our assumptions. In equation \ref{eq:Xi}, we see that for a fixed $T_e$ and $v_{\rm max}$, $M_{\rm ej}/\sqrt{f_V}$ is also fixed. The filling factor can therefore be understood as a factor that only affects the derived ejecta mass, and has no other effect on the radio light curve. If the filling factor were to decrease by an order of magnitude, it would decrease the derived ejecta mass by a factor of $\sim$3.

\begin{figure}[h]
\includegraphics[width=\columnwidth]{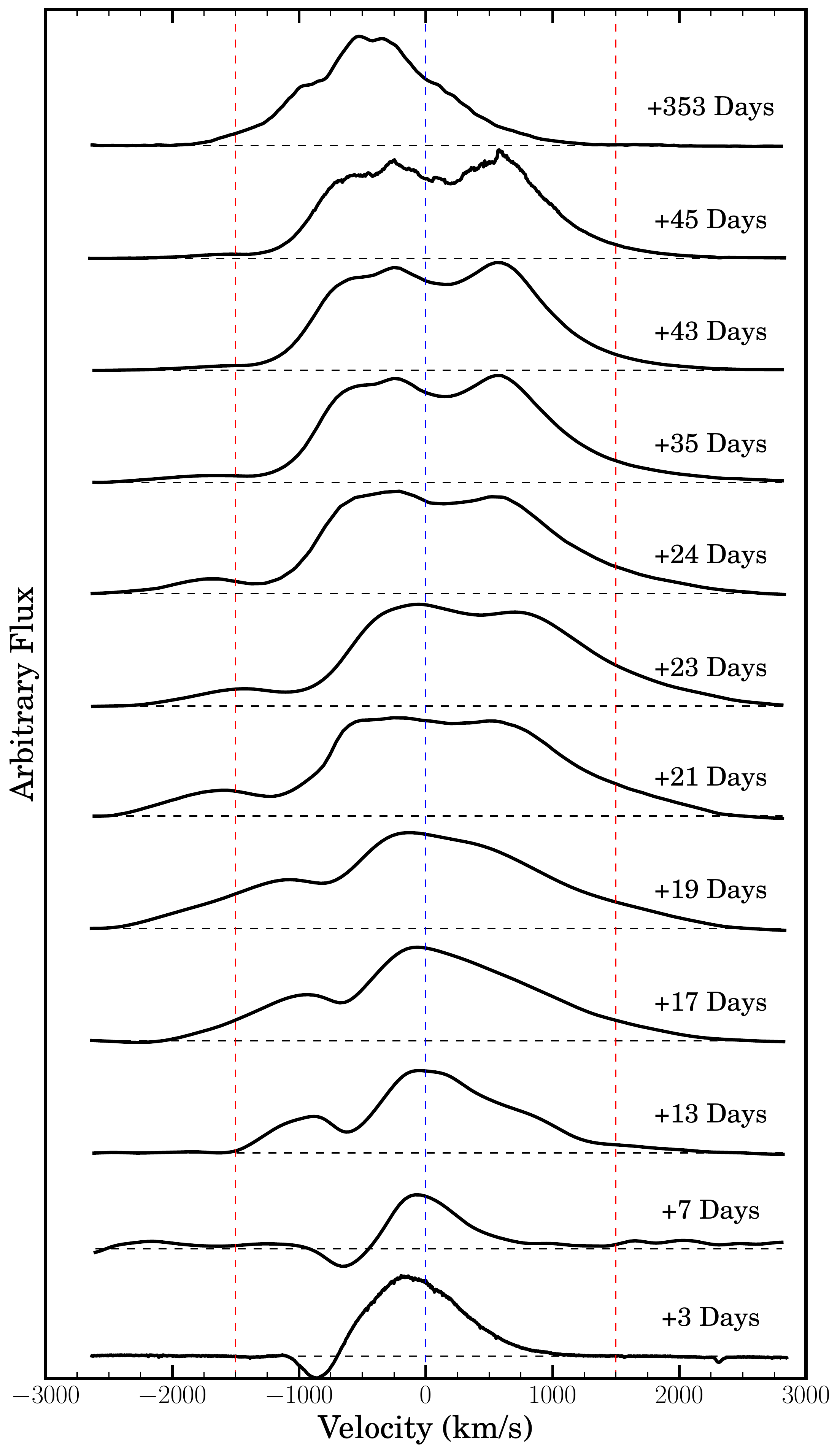}
\caption{\label{fig:HAlphaEvol}
Evolution of the H$\alpha$ line as a function of time. We take day 0 to be June 1 2012. All velocities have been corrected to the heliocentric frame. The blue dashed line indicates $v=0$ km s$^{-1}$, while the red dashed lines---used to help guide the eye---give $v=\pm 1500$ km s$^{-1}$. The $y-$axis is arbitrary flux; these relative flux values are not to scale. Note the expansion of the velocity profile starting sometime between day $+7$ and $+13$, and continuing until day $\sim +35$.
}\end{figure}

We now consider how ejecta mass depends on distance. Equation \ref{eq:Psi} implies that at least one of $D$, $v_{\rm max}$, and $T_e$ needs to be left free to vary in order to provide a suitable fit to a light curve. If we fix $D$ and instead let $v_{\rm max}$ vary, we find 
\begin{equation}
v_{\rm max} = D \Psi^{1/2} T_{e}^{-1/2}.
\end{equation}
(see Appendix A for a discussion of $\Psi$). We can fix $D$ at the minimum possible distance, $D = 6.5$ kpc \citep{2015ApJ...809..160F}, and maintain $T_e = 10^4$ K; then the implied maximum ejecta velocity is 1150 km s$^{-1}$ (consistent with the velocity of the P Cygni absorption trough in early spectra; Figure \ref{fig:HAlphaEvol}). The lack of observed velocities $>$2600 km s$^{-1}$ implies that V1324~Sco is not located further than 15 kpc away, unless its thermal ionized ejecta are somehow substantially cooler than $10^4$ K (which we consider very unlikely; e.g., \citealt{2015ApJ...803...76C}). A velocity of 1145 km s$^{-1}$ in turn implies an ejecta mass almost an order of magnitude lower, $2.3 \times 10^{-6}$ M$_{\odot}$.


It should be noted that, during the dust event, we expect some fraction of the nova ejecta to cool, recombine, and become neutral. Since neutral particles won't emit free-free emission---or, at least for atoms with significant dipole moments, they will emit significantly less free-free emission than ionized particles---we don't expect this mass to show up in the radio emission. However, the bulk of the second radio maximum occurs after the dust event, when the ionization of the gas should be increasing from a minimum around day $+$70 and approaching a photoionized equilibrium with temperature, $10^4$ K \citep{2015ApJ...803...76C}. 

Despite uncertainties, radio light curves remain one of the most robust ways to estimate the ejecta masses of classical novae \citep{Seaquist_Bode08}. We conclude that, given measured ejecta velocities in excess of 2000 km s$^{-1}$ and the lower limit on the distance, our measurements imply an ejecta mass for V1324~Sco of a few $\times 10^{-5}$ M$_{\odot}$.

\clearpage
\begin{turnpage}
\begin{center}
\begin{deluxetable}{cccccccccc}
\tabletypesize{\scriptsize}
\tablecaption{\label{tab:RadioObs}VLA Observations of V1324 Sco}
\tablehead{
\colhead{Julian } \vspace{-0.1cm} & \colhead{Date} &  &  & \colhead{4.5 GHz Flux \tablenotemark{a, b}} & \colhead{7.8 GHz Flux} & \colhead{13.3 GHz Flux}  & \colhead{17.4 GHz Flux} & \colhead{27.5 GHz Flux} & \colhead{36.5 GHz Flux} \\ \vspace{-0.1cm} 
& & $t-t_0$ & \colhead{Config.} & & & & & & \\ 
\colhead{($245000+$)} & \colhead{(UT)} & & & \colhead{(mJy)} & \colhead{(mJy)} & \colhead{(mJy)} & \colhead{(mJy)} & \colhead{(mJy)} & \colhead{(mJy)} 
}
\startdata
6104.3 & 6/26/2012 & 25.0   & B  & 0.165  $\pm$ 0.023 & 0.149  $\pm$ 0.021 & -- & -- & -- & -- \\
6106.1 & 6/28/2012 & 27.0   & B  & -- & -- & -- & -- & -- & $<$0.822 \\
6144.1 & 8/5/2012 & 65.0   & B  & 0.605  $\pm$ 0.115 & 2.074  $\pm$ 0.160 & -- & -- & -- & -- \\
6147.1 & 8/8/2012 & 68.0   & B  & 0.817  $\pm$ 0.124 & 2.720  $\pm$ 0.191 & -- & -- & -- & -- \\
6151.2 & 8/12/2012 & 72.0   & B  & 1.385  $\pm$ 0.100 & 2.803  $\pm$ 0.171 & -- & -- & 5.133  $\pm$ 0.636 & 5.647  $\pm$ 0.766 \\
6159.2 & 8/20/2012 & 80.0   & B  & -- & -- & 2.574  $\pm$ 0.307 & 2.690  $\pm$ 0.364 & 3.246  $\pm$ 0.489 & 2.639  $\pm$ 0.533 \\
6167.0 & 8/28/2012 & 88.0   & B  & 1.036  $\pm$ 0.092 & 1.128  $\pm$ 0.089 & -- & -- & -- & -- \\
6171.1 & 9/1/2012 & 92.0   & B  & -- & -- & 1.169  $\pm$ 0.144 & 1.250  $\pm$ 0.165 & 1.760  $\pm$ 0.261 & 2.370  $\pm$ 0.377 \\
6177.9 & 9/7/2012 & 98.0   & BnA  & 0.502  $\pm$ 0.055 & 0.431  $\pm$ 0.038 & -- & -- & -- & -- \\
6182.1 & 9/12/2012 & 103.0  & BnA  & -- & -- & 0.810  $\pm$ 0.128 & 1.080  $\pm$ 0.165 & 1.738  $\pm$ 0.290 & 2.126  $\pm$ 0.392 \\
6215.0 & 10/15/2012 & 136.0  & A  & $<$0.218 & 0.338  $\pm$ 0.049 & 0.843  $\pm$ 0.121 & 1.201  $\pm$ 0.174 & 2.085  $\pm$ 0.332 & 2.460  $\pm$ 0.415 \\
6243.8 & 11/12/2012 & 164.0  & A  & 0.228  $\pm$ 0.055 & 0.484  $\pm$ 0.050 & 1.297  $\pm$ 0.174 & 1.955  $\pm$ 0.256 & 4.153  $\pm$ 0.582 & 5.667  $\pm$ 0.838 \\
6297.6 & 1/5/2013 & 218.0  & A  & 0.353  $\pm$ 0.039 & 0.701  $\pm$ 0.051 & 1.699  $\pm$ 0.198 & 2.539  $\pm$ 0.301 & 4.240  $\pm$ 0.574 & 6.230  $\pm$ 0.983 \\
6357.5 & 3/6/2013 & 278.0  & D  & 0.761  $\pm$ 0.075 & 1.300  $\pm$ 0.121 & 2.877  $\pm$ 0.337 & 4.071  $\pm$ 0.469 & 5.660  $\pm$ 0.725 & 6.821  $\pm$ 0.917 \\
6402.3 & 4/20/2013 & 323.0  & D  & 0.963  $\pm$ 0.216 & 1.352  $\pm$ 0.153 & 2.810  $\pm$ 0.313 & 3.637  $\pm$ 0.403 & 4.677  $\pm$ 0.543 & 5.089  $\pm$ 0.605 \\
6445.4 & 6/2/2013 & 366.0  & DnC & 1.037  $\pm$ 0.186 & 1.507  $\pm$ 0.139 & 2.855  $\pm$ 0.322 & 3.283  $\pm$ 0.379 & 3.290  $\pm$ 0.429 & 3.680  $\pm$ 0.538 \\
6501.3 & 7/28/2013 & 422.0  & C  & 1.041  $\pm$ 0.105 & 1.446  $\pm$ 0.118 & 2.318  $\pm$ 0.270 & 2.337  $\pm$ 0.286 & 3.017  $\pm$ 0.659 & 3.160  $\pm$ 0.811 \\
6584.0 & 10/19/2013 & 505.0  & B  & 0.937  $\pm$ 0.084 & 1.440  $\pm$ 0.093 & 1.933  $\pm$ 0.218 & 1.773  $\pm$ 0.206 & 1.682  $\pm$ 0.222 & 1.486  $\pm$ 0.213 \\
6649.7 & 12/23/2013 & 570.0  & B  & 0.963  $\pm$ 0.106 & 1.120  $\pm$ 0.087 & 1.350  $\pm$ 0.182 & 1.350  $\pm$ 0.186 & 1.027  $\pm$ 0.205 & 1.098  $\pm$ 0.304 \\
6719.5 & 3/3/2014 & 640.0  & A  & 0.719  $\pm$ 0.068 & 0.841  $\pm$ 0.064 & 0.899  $\pm$ 0.114 & 0.843  $\pm$ 0.119 & 0.696  $\pm$ 0.143 & 0.529  $\pm$ 0.132 \\
7009.8 & 12/18/2014 & 930.0  & C  & 0.368  $\pm$ 0.058 & 0.293  $\pm$ 0.034 & 0.268  $\pm$ 0.044 & 0.244  $\pm$ 0.044 & $<$0.282 & $<$0.365 \\
\enddata
\vspace{-0.1in}
\tablecomments{Taking $t_0$ to be 2012 June 1}
\tablenotetext{a}{Detections are defined as flux $>5\sigma$. Non-detections are given as the $5\sigma$ upper limits.}
\tablenotetext{b}{If no observations were taken for a given frequency it is denoted by --.}
\end{deluxetable}
\end{center}
\end{turnpage}
\clearpage

\begin{figure*}[ht]
\includegraphics[width=\textwidth]{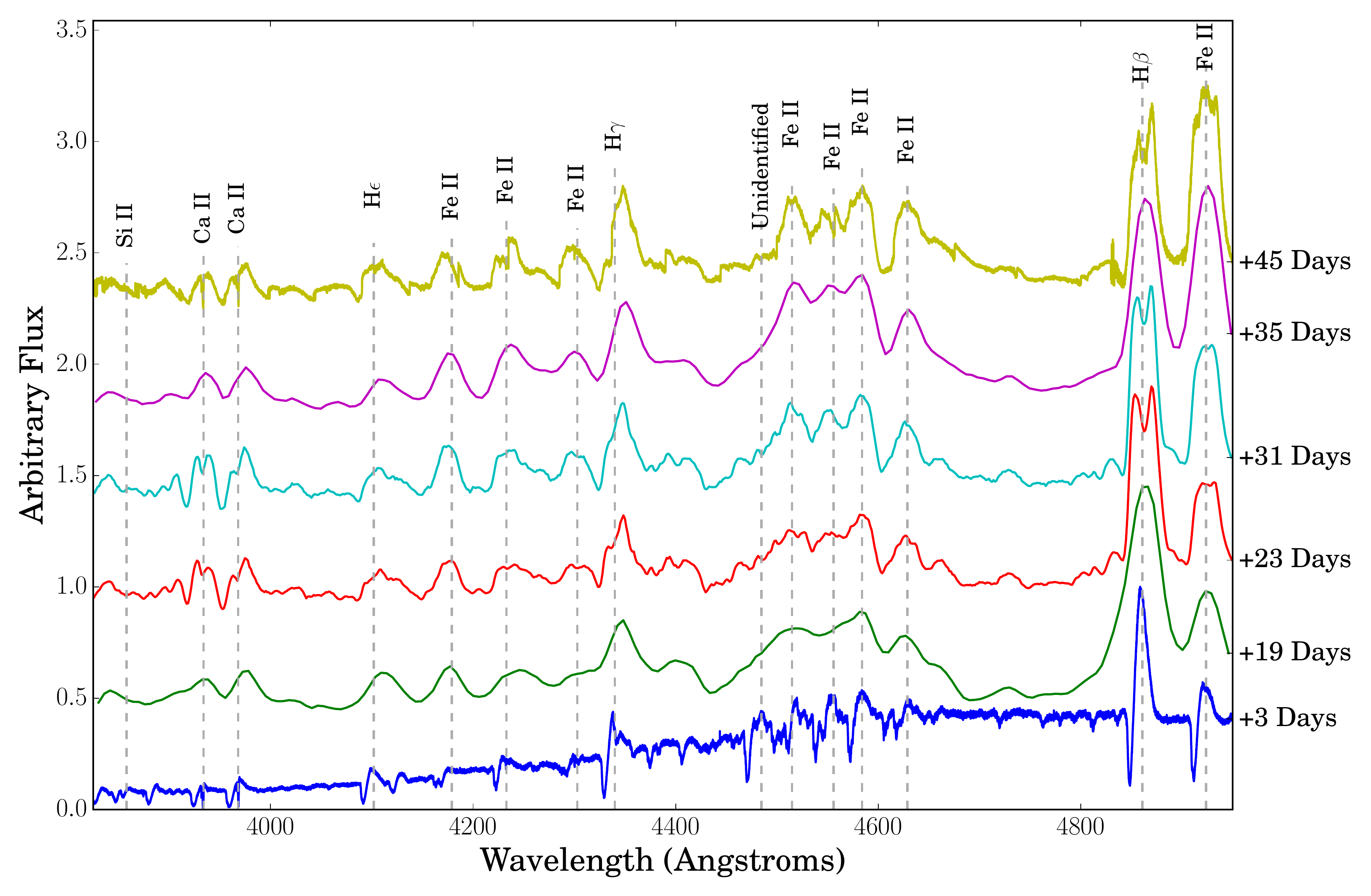}
\caption{\label{fig:SpectraBLUE}
Evolution of the blue ($3850-4950$~\AA) spectral region. All wavelengths have been corrected to the heliocentric frame. None of these spectra have been corrected for telluric features. See Table~\ref{tab:AllSpecObs} for details on the telescopes and instruments used for the different spectra. Note that GeV gamma-rays were observed during the second, third, and fourth spectra (days $+14-31$; \citealt{2014Sci...345..554A}).
}\end{figure*}

\begin{figure*}[ht]
\includegraphics[width=\textwidth]{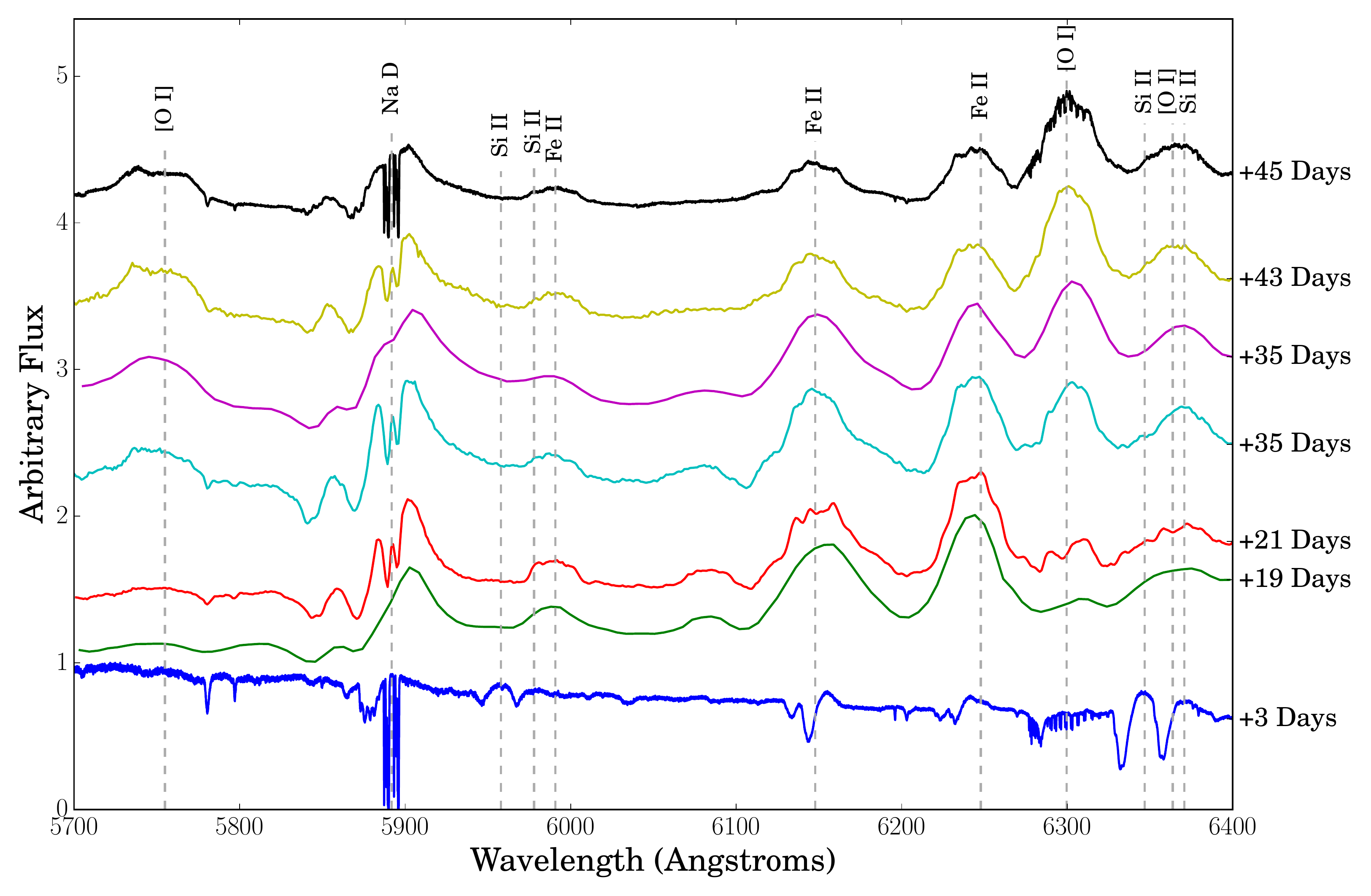}
\caption{\label{fig:SpectraREDL}
Evolution of the red ($5700-6400$~\AA) spectral region. All wavelengths have been corrected to the heliocentric frame. None of these spectra have been corrected for telluric features. The UVES spectrum taken on day $+3$ has contamination from telluric absorption lines between 6280~\AA\ and 6320~\AA\ . See Table~\ref{tab:AllSpecObs} for details on the telescopes and instruments used for the different spectra. Note that GeV gamma-rays were observed during the second and third spectra (days $+14-31$; \citealt{2014Sci...345..554A}).
}\end{figure*}

\begin{figure*}[ht]
\includegraphics[width=\textwidth]{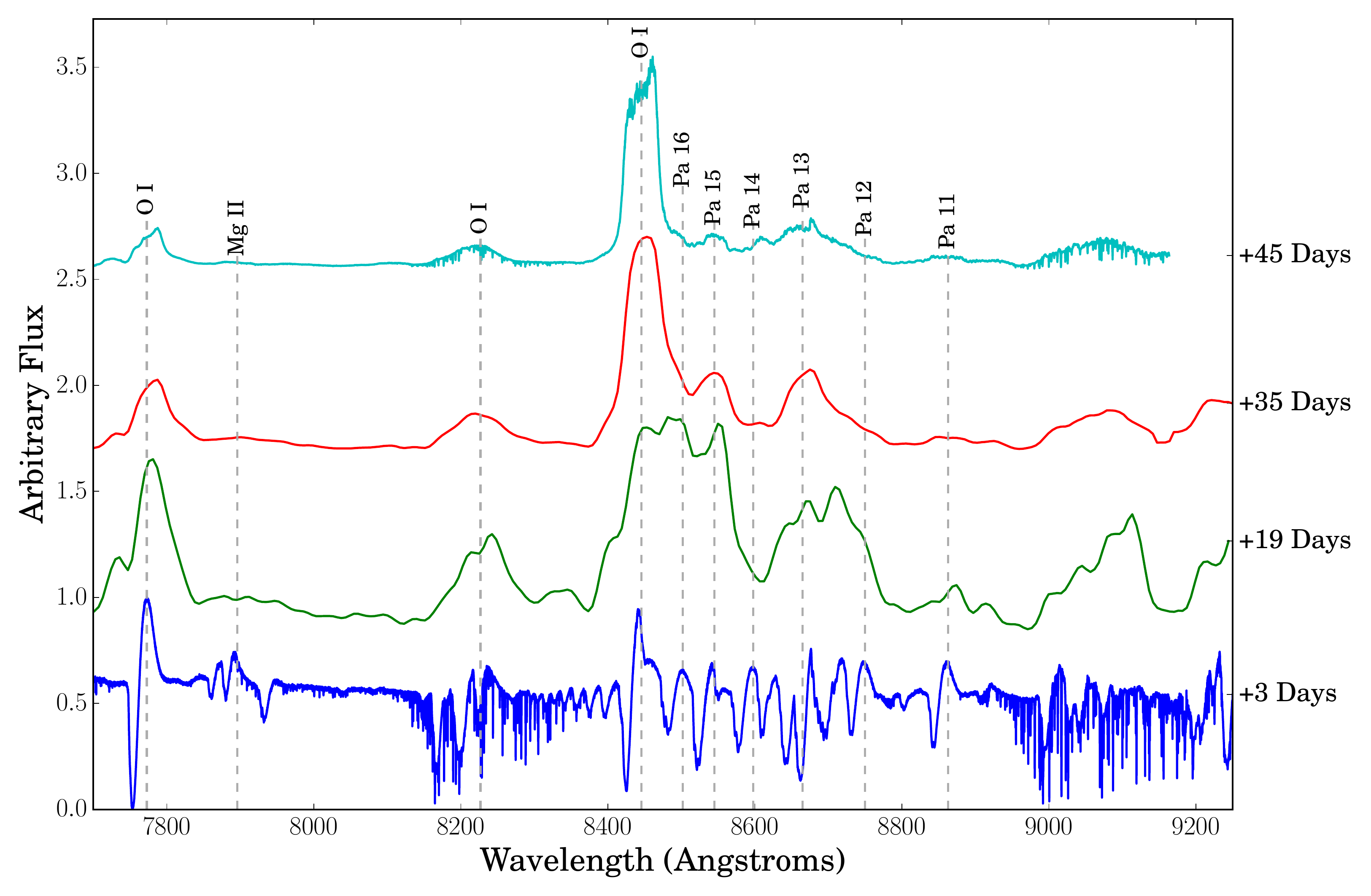}
\caption{\label{fig:SpectraREDU}
Evolution of the near-infrared ($7700-9250$~\AA) spectral region. All wavelengths have been corrected to the heliocentric frame. None of these spectra have been corrected for telluric features. The UVES spectrum taken on day $+3$ has prominent contamination from telluric absorption lines between 8200~\AA\ $-$ 8300~\AA\ and between 8900~\AA\ $-$ 9200~\AA . See Table~\ref{tab:AllSpecObs} for details on the telescopes and instruments used for the different spectra. Note that GeV gamma-rays were observed during the second spectrum (days $+14-31$; \citealt{2014Sci...345..554A}).
}\end{figure*}

\begin{figure}[h]
\includegraphics[width=\columnwidth]{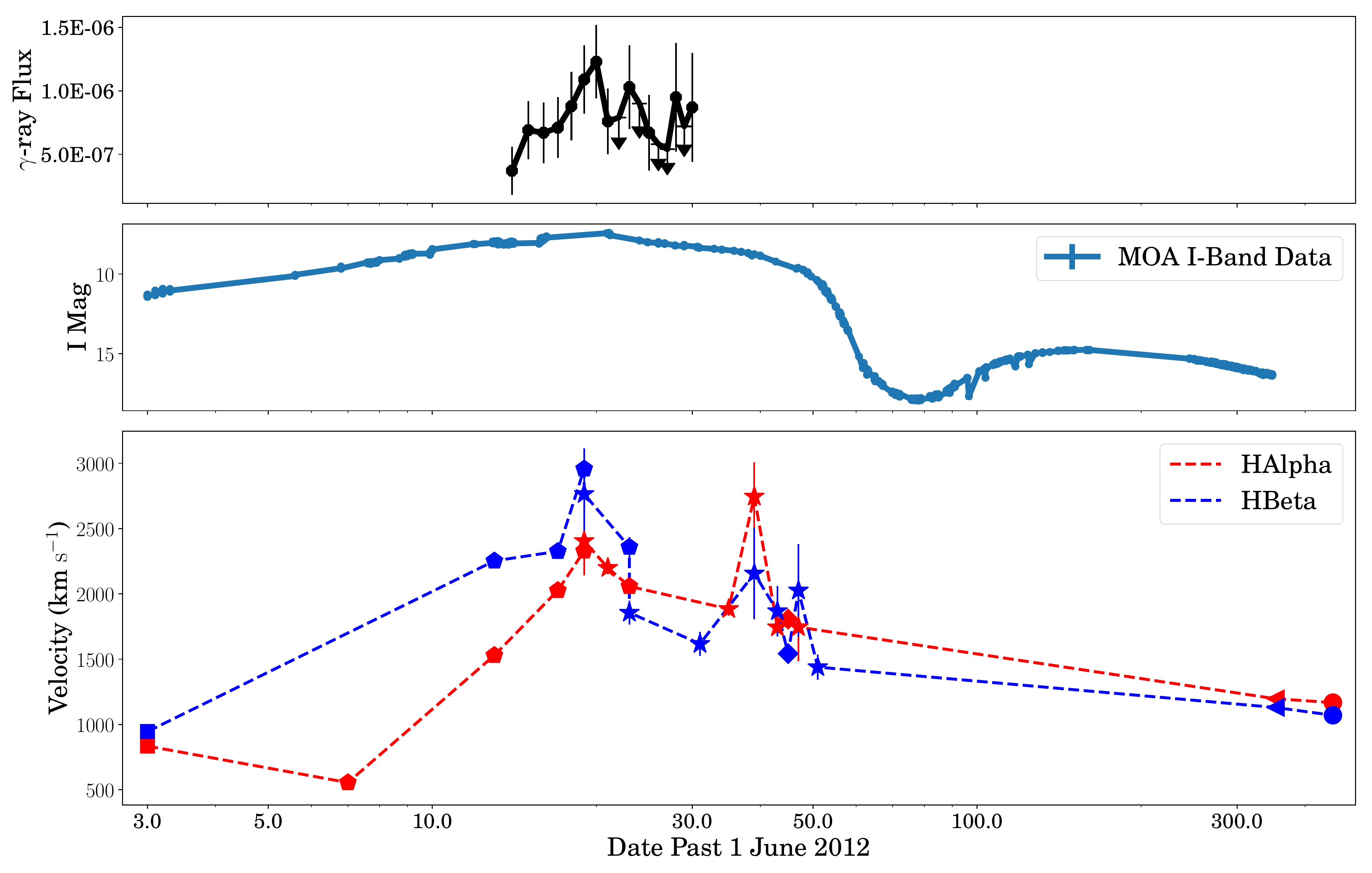}
\caption{\label{fig:BalmerLineVelocities}
Top Panel: The GeV gamma-ray light curve of V1324 Sco, as observed by \emph{Fermi}/LAT and published in \citep{2014Sci...345..554A}. Middle Panel: A subset of the $I$ band light curve in Figure~\ref{fig:OpticalLC}, given here for reference. Bottom Panel: Evolution of the velocities of both H$\alpha$ and H$\beta$. Both velocities are found by taking the Half-Width at Half Maximum HWHM of the spectral feature. This increase in velocity---coincident with the gamma-ray emission---is interpreted as being a signature of the shock interaction. 
}\end{figure}

\section{Optical Spectra}
Optical spectroscopy of novae are very rich and complex, but our primary goal for V1324~Sco is to understand the kinematics and energetics of the ejecta. Therefore, in this section we particularly focus on the gas kinematics and filling factor of the gas (which are crucial for estimating the ejecta mass from radio light curves; Section \ref{sec:RadioFitting}).

\label{sec:OpticalSpectra}
\subsection{Observations and Reduction}
All spectroscopic observations---including date, telescope, and observer---are listed in Table~\ref{tab:AllSpecObs}. Spectra are shown in Figures \ref{fig:SpectraBLUE}, \ref{fig:SpectraREDL}, and \ref{fig:SpectraREDU}. Note that all plots have been corrected to put spectra into the heliocentric frame.

\begin{center}
\begin{deluxetable*}{ccccccc}
\tabletypesize{\scriptsize}
\tablecaption{\label{tab:AllSpecObs} Optical Spectroscopic Observations}
\tablehead{
\colhead{UT Date}&
\colhead{$t-t_0$}&
\colhead{Observer}&
\colhead{Telescope}&
\colhead{Instrument}&
\colhead{Dispersion}&
\colhead{Wavelength Range} \\
& \colhead{(Days)} &&&& \colhead{($\angstrom$)} & \colhead{($\angstrom$)}
}
\startdata
2012 Jun 04.0  & $+3.0$ & Bensby 	& VLT & UVES & $0.02$ & $3700 - 9500$ \\
2012 Jun 08.5 & $+7.5$ & Bohlsen 	& Vixen VC200L & LISA & $0.5$ & $3800 - 8000$ \\
2012 Jun 14.5 & $+13.5$ & Bohlsen & Vixen VC200L	& LISA & $0.5$ & $3800 - 8000$	\\
2012 Jun 18.5 & $+17.5$ & Bohlsen 	& Vixen VC200L	& LISA  & $0.5$ & $3800 - 8000$ \\
2012 Jun 20.9 & $+19.9$ & Buil	& 0.28 meter Celestron & LISA  & $\sim 0.6 $ & $3700 - 7250$	\\
2012 Jun 21.2 & $+20.2$ & Walter	& SMARTS 1.5m & RC & $\sim 5.5$ & $3240 - 9500$ \\
2012 Jun 23.1 & $+22.1$ & Walter	& SMARTS 1.5m & RC & $\sim 1.0$ & $5620 - 6940$ \\
2012 Jun 24.9 & $+23.9$ & Buil	& 0.28 m Celestron & LISA & $6.2$ & $3700  - 7250$ \\
2012 Jun 25.1 & $+24.1$ & Walter	& SMARTS 1.5m & RC & $\sim 1.5$ & $3650 - 5420$ \\
2012 Jul 03.0 & $+32.0$ & Walter	& SMARTS 1.5m & RC & $\sim 1.5$ & $3650 - 5420$ \\
2012 Jul 07.1 & $+36.1$ & Walter	& SMARTS 1.5m & RC & $\sim 1.0$ & $5620 - 6940$ \\
2012 Jul 11.1 & $+40.1$ & Walter	& SMARTS 1.5m & RC & $\sim 5.5$ & $3240 - 9500$ \\
2012 Jul 15.0 & $+44.0$ & Walter	& SMARTS 1.5m & RC  & $\sim 1.0$ & $5620 - 6940$ \\
2012 Jul 16.1  & $+45.1$ & Chomiuk	& Clay Magellan & MIKE & $0.035$ & $3700 - 9200$ \\
2012 Jul 19.0& $+48.0$ & Walter	& SMARTS 1.5m & RC& $\sim 5.5$ & $3240 - 9500$ \\
2013 May 20.0 & $+353.0$ & Wagner	& LBT & MODS1 & $\sim 3.5$ & $3420 - 10000$ \\
2013 Aug 04.0 & $+450.0$ & Chomiuk	& SOAR & Goodman & $\sim 1.0$ & $3000 - 7000$	
\enddata
\end{deluxetable*}
\end{center}

The details of the data reduction for the UVES and MIKE data can be found in~\cite{2015ApJ...809..160F} and~\cite{2012PASP..124.1057W} for the RC Spectrograph data. The SOAR Goodman data were taken using a 400 l/mm grating centered at 5000~\AA, and were reduced using the standard procedure in IRAF \footnote{IRAF is distributed by the National Optical Astronomy Observatories, which are operated by the Association of Universities for Research in Astronomy, Inc., under cooperative agreement with the National Science Foundation. See \citet{Tody93}.} with optimal extraction and wavelength calibration using FeAr arcs. 

An optical spectrum was obtained on 2013 May 20.4 UT (day 353) using the 8.4~m Large Binocular Telescope (LBT) and Multi--Object Double Spectrograph (MODS1).  Observing conditions were photometric but the seeing as measured from two independent sources ranged from 1.8$-$1.9\arcsec\ at the start of the observation.  MODS1 utilized a 0\farcs8 entrance slit (so there was some loss of light at the entrance slit) and G400L (blue channel; 3200--5800 \AA) and G670L (red channel; 5800--10000 \AA) gratings giving a final dispersion of 0\farcs5 per pixel. The combined spectrum covers the range 3420--10000 \AA\ at a spectral resolution of 3.5 \AA. The spectra of quartz--halogen and HgNeArXe lamps enabled the removal of pixel--to--pixel and other flatfield variations in response and provided wavelength calibration respectively. Spectra of the spectrophotometric standard star BD+33 2642 were obtained to measure the instrumental response function and provide flux calibration of the V1324 Sco spectra. The spectra were reduced using a set of custom routines to remove the bias from the detectors and provide flatfield correction and then using IRAF for spectral extraction and wavelength and flux calibration.

In the case of the spectra taken by C. Buil and T. Bohlsen, both observers used a LISA spectrograph attached to commercially available telescopes of different sizes (0.28 meter Celestron for Buil; 0.22 meter Vixen VC200L for Bohlsen). More information about their observations can be found on their websites\footnote{\url{http://www.astrosurf.com/buil/index.htm}}$^{,}$\footnote{\url{http://users.northnet.com.au/~bohlsen/Nova/}}.

\subsection{Spectroscopic Evolution}
\label{sec:SpectraEvol}

As seen in  Figure~\ref{fig:HAlphaEvol}, there were strong P-Cygni absorption profiles starting at least as early as day $+3$. The H$\alpha$ emission component peaked at $\sim -180~\mathrm{km}~\mathrm{s}^{-1}$ on day $+3$, and had a FWHM of $\sim 800~\mathrm{km}~\mathrm{s}^{-1}$. 
The entirety of H$\alpha$, including both the emission feature and the P-Cygni absorption, extended out to $-1100~\mathrm{km}~\mathrm{s}^{-1}$ in the blue, or $900~\mathrm{km}~\mathrm{s}^{-1}$ from the line center. We take the P-Cygni absorption profile to be coming from the fastest material, meaning that---at this early time---the maximum expansion velocity was $\sim 900~\mathrm{km}~\mathrm{s}^{-1}$.
The second most prominent features in the early spectra---aside from the Hydrogen lines---are the \ion{Fe}{2} lines, all of which showed P-Cygni profiles. This is evident in Figures~\ref{fig:SpectraBLUE} and~\ref{fig:SpectraREDL}, which show the time evolution of the blue ($3850-4950$~\AA) and red ($5700-6400$~\AA) spectral regions, respectively. 

The spectrum obtained on day $+$13 shows the H$\alpha$ line profile clearly broadening (Figure \ref{fig:HAlphaEvol}). Note that this is also the time when the light curve flattens out, and stays at roughly constant brightness for the next month (Section \ref{sec:OpticalTimeline}). Between days $+$13--35, we see emission wings of the H$\alpha$ line expand to $\pm$2600 km s$^{-1}$ from the line center. We also see the P Cygni absorption trough move blueward during this time. We discuss the physical implications of this line broadening further in Section~\ref{sec:VelocityVariations}.

\begin{figure*}[ht]
\includegraphics[width=\textwidth]{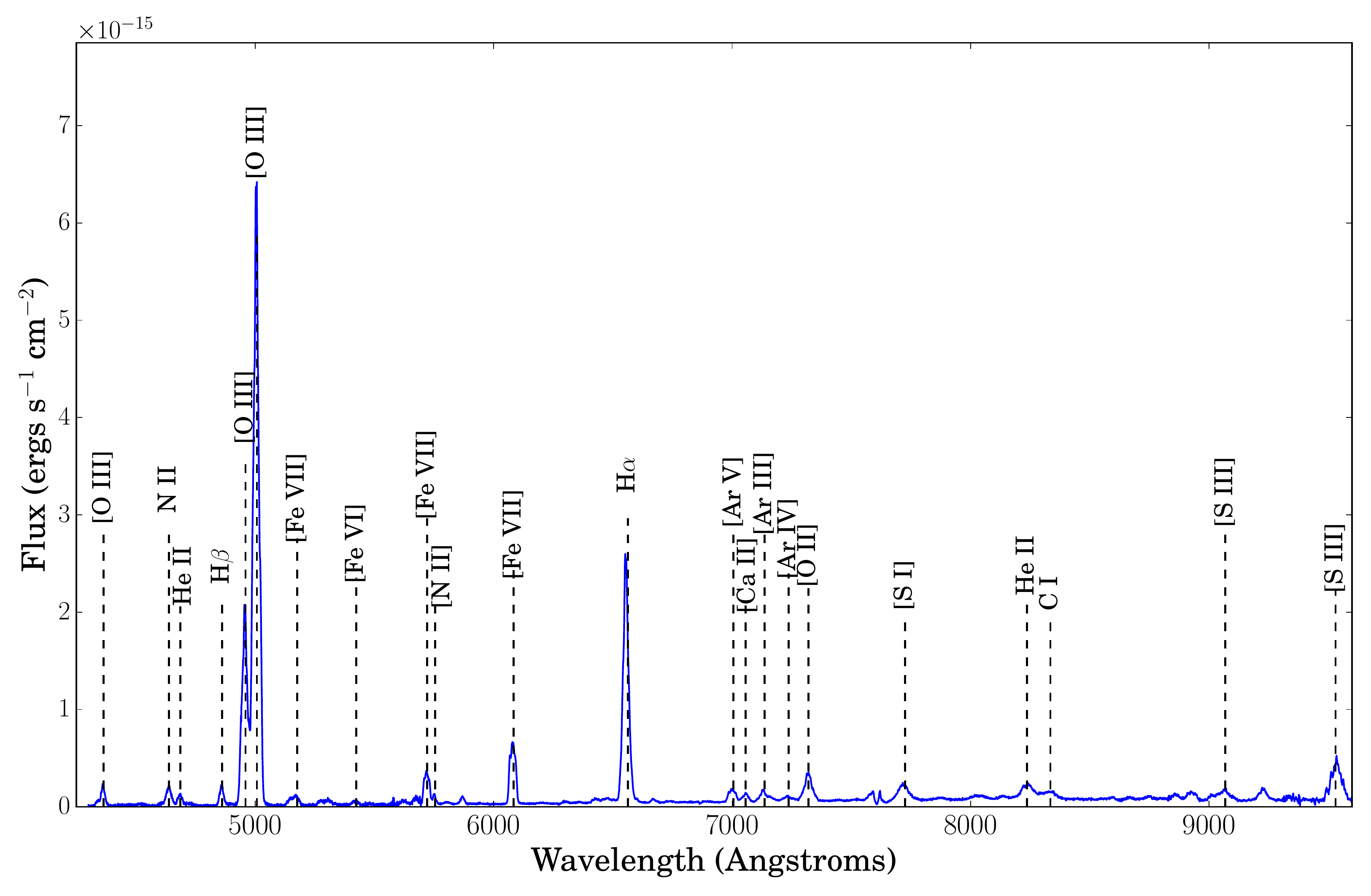}
\caption{\label{fig:SpectraNEBULAR}
Spectrum for V1324 Sco during the nebular phase taken on day $+353$, taken with the MODS1 instrument on the Large Binocular Telescope. See Table~\ref{tab:AllSpecObs} for further details on this spectrum. Lines were identified using the table provided in the appendix of~\cite{2012AJ....144...98W}.}
\end{figure*}

Just a few days after the Magellan MIKE spectrum (day $+45$), V1324 Sco underwent a massive dust dip lasting for $\sim 50$ days. Although the light curve did eventually rebound out of the dust dip, there was only a brief window of $\lesssim 25$ days before V1324~Sco went into solar conjunction. As a result our spectroscopic coverage did not pick back up until 20 May 2013---$355$ days after outburst---well into the nebular phase. As seen in Figure~\ref{fig:SpectraNEBULAR} the strongest lines in the nebular phase are the [\ion{O}{3}] lines at 5007 and 4959~\AA , followed by H$\alpha$ and [\ion{Fe}{7}] at 6084~\AA. 

\subsection{Discussion of Optical Spectra}
\label{sec:VelocityVariations}

V1324 Sco is a \ion{Fe}{2} type nova~\citep{1991ApJ...376..721W}, due to the prominence of the \ion{Fe}{2} spectral features---second only to the Balmer features---during optical maximum. The \ion{Fe}{2} type classification is common among \emph{D} type novae, including FH Ser, NQ Vul, and QV Vul (see~\citealt{2010AJ....140...34S} and references within). It is also notable that all \emph{Fermi}-detected novae to date have been of the \ion{Fe}{2} type---see V1369~Cen \citep{Izzo13}, V5668~Sgr \citep{Williams15}, V339~Del \citep{Tajitsu15}, and V5856 Sgr \citep{Luckas16, Rudy16}.

Looking at Figure~\ref{fig:HAlphaEvol}, it is clear that the Balmer line profile changes as a function of time. This type of line profile evolution is common amongst novae (\citealt{Gaposchkin57, McLaughlin60}; for some recent examples see e.g.,~\citealt{2014AJ....147..107S, 2014A&A...569A.112S, Shore16}).
The spectroscopic velocities for H$\alpha$ and H$\beta$ are plotted in Figure~\ref{fig:BalmerLineVelocities}, along with the photometric light curve for comparison purposes. Velocities quoted are half-width at half-maximum (HWHM) measured for H$\alpha$ and H$\beta$.  Because a number of the spectra taken by Walter et al.\ were either blue (3650--5420 \AA) or red (5620--6940 \AA), we chose to use both of these features to maximize the number of velocity measurements. The HWHM was measured by fitting a Gaussian profile to the emission lines using the IRAF routine \verb|splot|. Uncertainties in HWHM were found by adding in quadrature both the  uncertainty in the line measurement---found by measuring the line multiple times in \verb|splot|---and the (average) dispersion of the spectrum. 

In V1324~Sco, the width of the Balmer lines increases around the time that gamma rays are first detected (day $+14$). The HWHM velocity then varies, but stays at a large value ($\lesssim$1500 km s$^{-1}$) over the time period when gamma rays are observed (until day $+31$, shown as top panel in Figure \ref{fig:BalmerLineVelocities}). Another spike is seen in the velocity evolution around day $+40$, and then the velocity appears to decline as the nova transitions to its nebular phase.  

The profile evolution of the Balmer lines implies that there is relatively slow-moving material in the outer parts of the ejecta, surrounding faster internal material. This conclusion is common in studies of classical novae, and is by no means peculiar to V1324~Sco \citep[e.g.,][]{McLaughlin60, Friedjung66, OBrien94}. In V1324 Sco,
P Cygni profiles apparent in early spectra imply that the outer, slow component is expelled at $\sim$1,000 km s$^{-1}$ (day $+3$ in Figure \ref{fig:BalmerLineVelocities}). Over the next couple weeks, a faster component of ejecta becomes visible, reaching velocities of $\sim$2,600 km s$^{-1}$. The delayed appearance of this fast component implies that it must be internal to the slow component (possibly because it is launched later, or over a longer period of time). Inevitably, the internal fast component will catch up with the external slow component, producing shocks (and gamma rays; see Section \ref{sec:DiscussionGammaRayNova}). Therefore, from the line profile evolution of V1324~Sco, we estimate that the differential velocity in the shock is $\sim$1,600 km s$^{-1}$.

It is unclear if the temporal correspondence between the broadening of the optical emission lines and the appearance of gamma rays is meaningful or coincidental. The optical emission lines of novae typically broaden in the weeks following optical maximum, as they transition from the principal line profile to show diffuse-enhanced line systems \citep{McLaughlin60}. It is possible that the fast, internal component is present in the nova practically since the start of outburst, but only becomes visible as the outer parts of the ejecta expand and drop in optical depth. However, the temporal coincidence between optical line broadening and gamma-ray turn-on is striking, and could hint that the fast component is not launched until $\sim$13 days into the outburst. Similar evolution can be seen in the H$\alpha$ profile of another gamma ray nova, V339 Del. Figure 4 of~\cite{2014A&A...569A.112S} show that the wings of the H$\alpha$ profile began to increase on 2013 August 18 (date of the first gamma-ray detection).

We can also use the spectroscopic observations to determine properties of the ejecta density in V1324~Sco. Specifically, we use the late-time (nebular) spectroscopy to measure density inhomogeneities (i.e., clumpiness) in the ejecta, which we parameterize in terms of the volume filling factor ($f_V$). Such inhomogeneities must be taken into account in order to get a proper mass estimate, and we incorporate the filling factor in our radio ejecta mass derivation in Section \ref{sec:RadioFitting}.
For detailed calculations of V1324~Sco's filling factor, see Appendix B. We use measurements of the [\ion{O}{3}] lines to find a filling factor of  $f_V = \FillingFactor~$. This is similar to the filling factor measured in the gamma-ray detected nova V339~Del ($f_V = 0.07-0.2$; \citealt{Shore16}). 

We also use the \ion{O}{1} lines measured on day $+45$---permitted transitions at 7774~\AA\ and 8446~\AA, and the forbidden transition at 6300~\AA---to constrain the column density (for at least some portions) of the ejecta (\citealt{1995ApJ...439..346K, 2012AJ....144...98W}; see Appendix C for the detailed calculations). If we assume a temperature of $T_e = 10,000$ K, the \ion{O}{1} ratios are consistent with density $\log (N_e/[\mathrm{cm}^{-3}]) > 10$. Assuming that the density scales like $t^{-3}$, we would expect the density to be a factor of $\sim 10$ times greater during the first X-ray observation (day $+$21) than it was on day $+45$. Combined with the fact that we expect the ejecta to have expanded to $\sim$ a few $\times 10^{14}$ cm, we derive a column density $\gtrsim 10^{24}$ cm$^{-3}$. As discussed in Section~\ref{sec:XRay}, such a high column density can explain the lack of hard X-ray emission.

\section{X-ray Data}
\label{sec:XRay}
Multiple X-ray observations were made using the \emph{Swift} X-Ray Telescope (XRT), all of them yielding non-detections. Non-detections span days $+21$ to $+520$, and include some observations coincident with the \emph{Fermi}/LAT detection of gamma rays. We present the X-ray limits obtained from the \emph{Swift} observations in Table~\ref{tab:allxray}. The quoted count rates are $3\sigma$ upper limits, derived using the Bayesian upper limit method outlined in~\cite{1991ApJ...374..344K}. The count rates were converted into luminosities assuming emission from a thermal plasma with characteristic temperature 1 keV and a distance of~\AssumedDistance\ (which is the lower limit derived in~\citealt{2015ApJ...809..160F}). Luminosity limits are quoted over the range $0.3-10$ keV, and only correct for absorption by the ISM, assuming a column density of $8\times 10^{21}~\mathrm{cm}^{-2}$. The column density was derived using the reddening values of~\cite{2015ApJ...809..160F} and the relationship of~\cite{2009MNRAS.400.2050G}. Note that these limits were also used in the analysis of~\cite{2014MNRAS.442..713M}.

X-rays from novae are often divided into two distinct components: optically-thick thermal X-rays from the hot white dwarf (i.e., super-soft source) and optically-thin harder thermal X-rays from shocked plasma \citep{Krautter08}. Recently, it has been proposed that non-thermal hard X-rays may also be present in novae, driven by the same  population of relativistic particles that produce the gamma rays \citep{2016arXiv161104532V}. The X-ray non-detection of V1324~Sco is especially noteworthy given that the high gamma-ray luminosity should imply a strong shock which, in turn, should generate a significant amount of hard X-rays~\citep{2001ApJ...551.1024M}. As discussed in~\cite{2016arXiv161104532V}, this apparent contradiction can be explained by either the presence of high densities behind the radiative shock---due to Coulomb collisions sapping energy from what would otherwise be X-ray emitting particles---or by bound-free (photoelectric) absorption or inelastic Compton downscattering if there is a large column of material ($\gtrsim 10^{25}$ cm$^{-2}$) ahead of the shock. In Appendix C, we use oxygen line ratios to show such high column densities are plausible.

Note that, along with the peculiar lack of hard X-rays from non-thermal particle acceleration, there was also a lack of soft X-rays, which are often seen at later times as the ejecta clear away and reveal the central hot white dwarf (e.g.,~\citealt{Schwarz11}). However, V1324 Sco was both distant ($\geq 6.5$ kpc; \citealt{2015ApJ...809..160F}) and suffered a large absorbing column density. The only other nova given in~\cite{Schwarz11} with both of these characteristics is V1663 Aql, a nova that was also never detected as a super-soft source. Another possible explanation for the lack of soft X-rays from V1324~Sco is that it occurred on a low-mass white dwarf, and the white dwarf photosphere was never hot enough to emit X-rays, instead peaking in the UV band \citep{Sala05, Wolf13}.
  
The X-ray behavior of V1324~Sco is consistent with other \emph{D} class nova. We know from recent \emph{D} class novae that it is the norm---rather than the exception---for dusty novae to go undetected in X-rays. As discussed in~\cite{Schwarz11}, only one \emph{D} class nova has been detected in hard X-rays: V1280~Sco, although this detection was $\sim 800$ days  after the beginning of the nova event.  Although not considered a \emph{D} class nova,~\cite{Schwarz11} makes the case that V2362~Cyg is another dusty nova that has been detected in hard X-rays. A further two marginally dusty novae were detected as super-soft-sources (V2467~Cyg and V574~Pup, see~\citealt{Schwarz11} and references within). Note that these two sources showed little to no change in their optical light; the presence of dust was only determined due to a modest increase in IR flux. On the other hand, six other dusty novae---including V1324~Sco---were observed but not detected in X-rays (V1324~Sco, V2676~Oph, V2361~Cyg, V1065~Cen, V2615~Oph, and V5579~Sgr; again, see~\citealt{Schwarz11}). This lack of X-ray emission in dusty novae could be explained if dust is a signature of radiative shocks and cold, dense shells \citep{2016arXiv161002401D}, which would absorb the majority of X-rays and re-emit them at optical wavelengths \citep{2014MNRAS.442..713M, 2015MNRAS.450.2739M}.

\begin{deluxetable}{lcccr}
\tabletypesize{\scriptsize}
\tablecolumns{4} 
\tablecaption{\label{tab:allxray} X-ray Upper Limits from \emph{Swift} XRT}
\tablehead{
\colhead{Date}  & \colhead{$t-t_0$} &  \colhead{Count Rate\tablenotemark{a}} & \colhead{Luminosity\tablenotemark{a}\tablenotemark{b}} & \\
\colhead{(UT)}& \colhead{(Days)} & \colhead{(s$^{-1}$)}& \colhead{(ergs s$^{-1}$})&
}
\startdata
2012 Jun 22  & $+21$ & $<$0.0031 &	$<$1.67E+33	\\
2012 Jun 27 & $+26$ & $<$0.0054 &	$<$2.91E+33	\\
2012 Jun 28 & $+27$ & $<$0.0151 &	$<$8.11E+33	\\
2012 Jul 4	&	$+33$ &	$<$0.0038 &	$<$2.04E+33	\\
2012 Jul 10	&	$+39$ &	$<$0.012	& $<$6.44E+33\\
2012 Jul 13	&	$+42$ &	$<$0.0055	& $<$2.96E+33\\
2012 Aug 14 & $+74$	& $<$0.0031	& $<$1.66E+33	\\
2012 Oct 16 & $+137$	& $<$0.0023	& $<$1.23E+33\\
2013 May 22 & $+355$ &	$<$0.003	& $<$1.61E+33\\
2013 Nov 3 &	$+520$	&	$<$0.0037	& $<$1.99E+33\\
\enddata
\tablenotetext{a}{$3\sigma$ Upper limits}
\tablenotetext{b}{Note that this is based on a distance lower bound of $6.5$ kpc. If the distance is greater, than the luminosity would also be greater.}
\end{deluxetable}

\section{Discussion}
\label{sec:Discussion}

\subsection{V1324 Sco: A Classical Nova}
\label{sec:DiscussionNormalNova}
In this section, we argue that all of the non-gamma-ray observational signatures of V1324 Sco are typical of classical novae, in the sense that all observational features have been seen in previous novae.

This point is relevant as there has been some discussion in the community that gamma-ray luminous systems like V1324~Sco may not be classical novae at all, but may instead belong to the class of intermediate-luminosity transients often called luminous red novae (LRN; e.g., \citealt{Blagorodnova17}). 
A luminous red nova, observationally, appears with persistently redder colors than a classical nova and luminosities that range from slightly fainter than classical novae to several magnitudes more luminous (e.g.,~\citealt{Kimeswenger02, Smith16}). The physical interpretation of the LRN optical/IR outburst is the merger of a close binary \citep{2011A&A...528A.114T, 2013Sci...339..433I,MacLeod17}. V1309~Sco is one of the canonical and best-studied LRNs, and its optical light curve is similar to V1324 Sco. Both have an initial, slow, monotonic rise, both have a flattening of the optical light curve near peak, and both have a significant dust event \citep{2011A&A...528A.114T}. 

We find, however, that a luminous red nova does not fit with the other observational characteristics of V1324~Sco. Unlike in LRN, the spectra of V1324~Sco evolve to show higher ionization species in the months following outburst (Figure \ref{fig:SpectraNEBULAR}). Additionally, V1324~Sco's ejecta velocities of a few thousand km s$^{-1}$ would be unusually high for a luminous red nova, which typically show velocities of a few hundred km s$^{-1}$ \citep{Munari02, Mason10}.

In addition, from the radio light curve of V1324~Sco, we estimate an ejecta mass of a few $\times 10^{-5} M_{\odot}$ (see Section \ref{sec:RadioFitting}). This is at least three orders of magnitude smaller than what is expected for LRN events, which are thought to expel a significant fraction of a solar mass \citep{2013Sci...339..433I, MacLeod17}. For the radio light curve to be consistent with $>10^{-1}$ M$_{\odot}$, V1324~Sco would need to be much further away than 15 kpc and expanding at a velocity $<<$1000 km s$^{-1}$. Such low velocities are implausible given the observed optical line profiles of V1324~Sco (Figure \ref{fig:HAlphaEvol}). In addition, when the ejecta mass is combined with the ejecta velocity, we find that the kinetic energy of the outburst of V1324~Sco was $10^{44}-10^{45}$ erg, typical for a classical nova. LRN, on the other hand, have kinetic energies $\sim 10^{47}-10^{48}$ erg (e.g.~\citep{Nandez14, MacLeod17,Metzger_Pejcha17}).

\begin{figure}[h]
\includegraphics[width=\columnwidth]{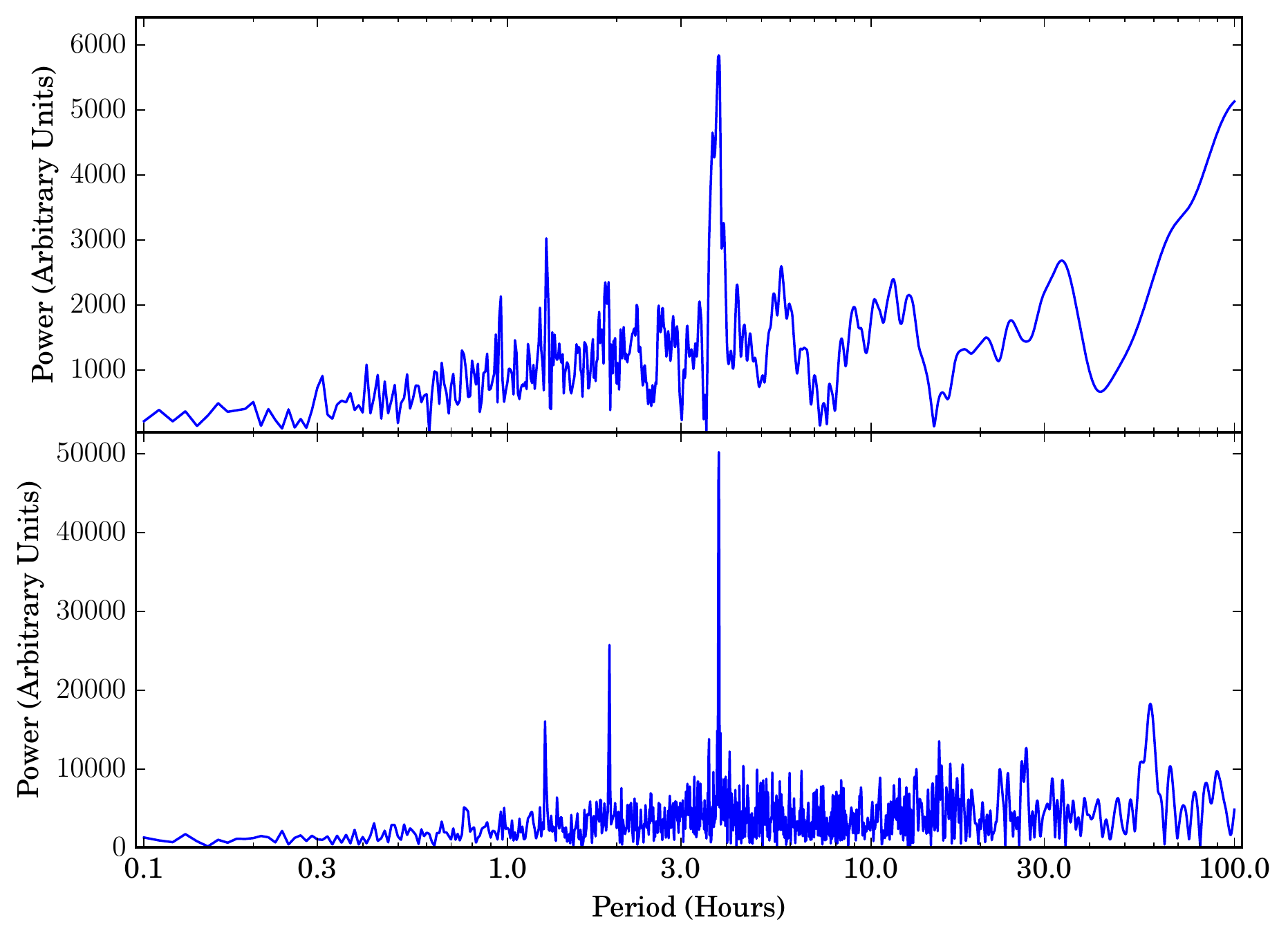}
\caption{\label{fig:PeriodoGrams}
Top Panel: Periodogram made using all of the $5\sigma$ detections in the MOA data taken before 2012 June 1 (i.e. during the Initial Rise phase). Bottom Panel: Periodogram made using all of the $5\sigma$ detections in the MOA data taken after 2012 October 1 (i.e. during the power law decline phase). Both show clear detections of a period at 3.8 hours, ruling out the possibility that V1324 Sco was a Luminous Red Nova. The level of noise is higher in the top plot because the initial rise phase lasted $<100$ days and had substantially fewer 5$\sigma$ detections than the much longer lasting power law decline phase (316 compared to 3677).}\end{figure}

In addition, we detect a photometric modulation in the optical light curve during the power law decline phase, which probably reflects the orbital period. We used the MOA data set, which had the best sampling, as well as the highest cadence;  we limited the data set to $>5\sigma$ detections. The periodic modulation was measured using the Lomb-Scargle algorithm in the Python scientific library \emph{SciPy}. We found it to be 3.8 hours, consistent with that observed in the precursor rise (Figure \ref{fig:PeriodoGrams}).
Assuming this periodicity reflects the underlying host binary, the detection of such modulation both before and after outburst implies that the binary was not destroyed in the nova event. Contrast this observation with measurements made for V1309~Sco, where \citet{2011A&A...528A.114T} watched the period dramatically decrease in the lead up to outburst, and then all periodic modulation disappeared. We therefore conclude that V1324 Sco is a classical nova, with host properties, ejecta mass, and kinetic energy consistent with other novae.

\subsection{V1324 Sco: The Most Gamma-Ray Luminous Nova to Date}
\label{sec:DiscussionGammaRayNova}
Having established in the previous section that V1324 Sco is a classical nova, we now discuss why this nova had such a high gamma-ray luminosity. As can be see in \citet{2016ApJ...826..142C}, V1324~Sco is the most gamma-ray luminous classical nova discovered to date. Note, however, that Cheung et al.\ assume a distance of 4.5 kpc to V1324~Sco, while \citet{2015ApJ...809..160F} showed that its distance is substantially further ($>$6.5 kpc). Therefore, V1324~Sco is even more gamma-ray luminous than presented in \citet{2016ApJ...826..142C}, registering at $L_{\gamma} \gtrsim 2 \times 10^{36}$ erg s$^{-1}$.

In V1324~Sco, there is strong observational evidence for shock interaction, both from the gamma rays and from the radio (in particular, the initial maximum; Section \ref{sec:FirstRadioBump}). We present here a simplified model of the ejecta that can explain the gamma-ray emission. 

We can imagine the ejecta as being composed of two parts: a slow initial component and a fast secondary component. When these two components meet, there will be both a forward and reverse shock, and it is these shocks that will power the gamma-ray emission. We further assume that these shocks are radiative, so we expect there to be a layer of cold material between the forward and reverse shocks. This analysis is based on the models of~\cite{2014MNRAS.442..713M} and \cite{2016MNRAS.463..394V}. 

Initially in the outburst, a slow component is expelled. We model it to be an impulsive ejection--- defined by \cite{2014MNRAS.442..713M} to have the density profile: 
\begin{equation}
n_{\rm s}(r) = \left(\frac{\dot{M}_{\rm s}}{4\pi f_{\Omega} v_{\rm s} r^{2}\mu m_p}\right)\exp\left[-\frac{r}{R_{\rm s}}\right],
\end{equation}
where $\dot{M}_{\rm s}$ is the slow component mass loss rate, $\mu$ is the mean molecular weight, $f_{\Omega}$ is the solid angle fraction that is subtended by the slow component, $v_{\rm s}$ is the velocity of the slow component, and $R_{\rm s}$ is the radius of the slow component ($R_{\rm s} = v_{\rm s} t_{\rm s}$, where $t_{\rm s}$ is the time since the slow component was ejected).

Thereafter, a faster component is expelled in a wind-like process. 
The fast component's density profile is given by
\begin{equation}
n_{\rm f}(r) \approx \frac{\dot{M}_{\rm f}}{4\pi v_{\rm f} r^{2}\mu m_p},
\end{equation}
where $\dot{M}_{\rm f}$ is the fast component mass loss rate, $v_{\rm f}$ is the velocity of the fast component. For V1324~Sco, we take $v_{\rm s}$ = 1000 km s$^{-1}$ and $v_{\rm f}$ = 2600 km s$^{-1}$ (Section \ref{sec:VelocityVariations}). 

The fast wind then impacts the slow component and shock interaction ensues. Assuming that the shock is radiative \citep{2014MNRAS.442..713M}, there will be a cold layer of material between the forward and reverse shocks (note that this is also the region where dust will form; \citealt{2016arXiv161002401D}). We denote the mass of this cold shell as $M_{\rm shell}$, which grows in time as
\begin{equation}
\frac{dM_{\rm shell}}{dt} = f_{\Omega}\dot{M}_{\rm f}\left(\frac{v_{\rm f} - v_{\rm shell}}{v_{\rm f}}\right) + \dot{M}_{\rm s} \left(\frac{v_{\rm shell}-v_s}{v_s}\right),
\end{equation}
while the momentum grows as 
\begin{equation}
\frac{d}{dt}\left(M_{\rm shell}v_{\rm shell}\right) = f_{\Omega}\dot{M_{\rm f}}(v_{\rm f}-v_{\rm shell}) + \dot{M}_{\rm s} (v_{\rm shell}-v_{\rm s}),
\end{equation}
where $\dot{M}_{\rm s}$ is evaluated at radius $r$ according to its value a time $t \approx r/v_{\rm s}$ before the onset of the fast wind.  A steady state solution (i.e. $dv_{\rm shell}/dt = 0$) is soon reached, wherein the shell gains most of its momentum from the fast wind and most of its mass by sweeping up the slow shell.  In this limit $v_{\rm f} \gg v_{\rm s}$ and $\dot{M}_{\rm f} \lesssim \dot{M}_{\rm s}f_{\Omega}^{-1}$ we find that 
\begin{equation}
\frac{v_{\rm shell}}{v_{\rm s}} \approx \left(\frac{\dot{M}_{\rm f} v_{\rm f}f_{\Omega}}{\dot{M}_{\rm s} v_{\rm s}}\right)^{1/2}.
\label{eq:vshell}
\end{equation}
The radius of the shell and the accompanying shock increases with time approximately as $R_{\rm shell} \approx v_{\rm shell}t$ as it passes through the slow ejecta.

The power dissipated by the shocks is determined by the number of thermal particles swept up by the shock, which can be expressed as
\begin{eqnarray}
\dot{E}_r &=& \frac{9}{32} f_{\Omega} \frac{\dot{M}_{\rm f}}{v_{\rm f}} (v_{\rm f} - v_{\rm shell})^3 \\
\dot{E}_f &=& \frac{9}{32} \frac{\dot{M}_{\rm s}}{v_{\rm s}} (v_{\rm shell} - v_{\rm s})^3. 
\end{eqnarray}
where $E_r$ and $E_f$ are the power dissipated at the reverse and forward shocks, respectively.
Since usually $v_{\rm shell} << v_{\rm f}$, the shock power (and gamma-ray luminosity) will be dominated by the reverse shock,
\begin{equation}
L_{\gamma} \propto \dot{E}_r = \frac{9}{32} f_{\Omega} \dot{M}_{\rm f} (v_{\rm f}-v_{\rm shell})^2 \approx  \frac{9}{32} f_{\Omega}\dot{M}_{\rm f} v_{\rm f}^2. 
\label{eq:Lgamma}
\end{equation}

To determine the amount of time that the gamma-ray emission will persist---hereafter referred to as $t_{\gamma}$---we need to find the amount of time it will take for the shock to cross the initial slow component, i.e. $R_{\rm shell} \approx v_{\rm shell}t_{\gamma} = R_{\rm s} = v_{\rm s} t_{\rm s}$.  Rewriting this using our expression for the shell/shock velocity (eq.~\ref{eq:vshell}) we find
\begin{equation}
t_{\gamma} = \frac{R_{\rm s}}{v_{\rm shell}} = \left(\frac{\dot{M}_{s}v_{\rm s}}{\dot{M}_{\rm f}v_{\rm f}f_{\Omega}}\right)^{1/2}t_{\rm s}.
\end{equation}
Using equation (\ref{eq:Lgamma}) we find an approximate proportionality
\begin{equation}
t_{\gamma} \propto \left(\frac{\dot{M}_{\rm s}v_{\rm s}v_{\rm f}}{L_{\gamma}}\right)^{1/2}t_{\rm s}
\end{equation}
From the above, we see that increasing either $\dot{M}_{\rm f}$ or $v_{\rm f}$ will increase $L_{\gamma}$ while generally decreasing $t_{\gamma}$ (for fixed values of $\dot{M}_{\rm s}$ and $v_{\rm s}$). Such an inverse relationship between the gamma-ray luminosity and the duration of the gamma-ray emission has been observed by~\cite{2016ApJ...826..142C}. 


The gamma-ray luminosity is proportional to the square of the  relative velocity between the fast component and the central shell (Equation~\ref{eq:Lgamma}), while $v_{\rm shell}$ is itself proportional to the velocity of the slow component.
Therefore, one possibility is that the differential velocity between the fast and slow components is unusually large in V1324~Sco.  We crudely estimate the differential velocity as $v_{\rm f} - v_{\rm s}$ = $(2600 - 1000)$ km s$^{-1}$ = 1600 km s$^{-1}$. It is still early to compare this quantity with many of the other \emph{Fermi}-detected novae, but sufficient studies have been published to consider V959~Mon and V339~Del. 

Optical spectroscopy of V339~Del has been studied by \citet{Shore16}; from their Figure 2, we estimate that a slow component is visible around day $+$5 with a velocity $v_{\rm s} \approx$ 1400 km s$^{-1}$. Later on, around day $+$22, a fast  wing appears on the H$\gamma$ profile extending out to $v_{\rm f} \approx$ 1900 km s$^{-1}$. In this case, both the fast component and the differential velocity are slower than in V1324~Sco, which might explain why V339~Del is a factor of $\gtrsim$4 less luminous in gamma rays (\citealt{2016ApJ...826..142C}; after taking into account the distance limit on V1324~Sco from \citealt{2015ApJ...809..160F}).

A direct comparison between V1324~Sco and V959~Mon, using the evolution of optical spectra, is impossible due to the fact that V959~Mon was in solar conjunction for the first months of its outburst. We can, however, use a combination of nebular spectroscopy and imaging of the nova ejecta to infer the velocities of the slow and fast component. \citet{Ribeiro13} model the nebular spectrum as expanding bipolar ejecta, and find a maximum velocity of 2400 km s$^{-1}$ (rather similar to V1324~Sco). \citet{Linford15} imaged the expanding ejecta at multiple times and frequencies using the VLA, and combining it with Ribeiro et al.'s model, found that the slow component expands at 480 km s$^{-1}$. According to this analysis, the differential velocity between fast and slow components in V959~Mon should be even larger than in V1324~Sco, but V959~Mon is a factor of $\gtrsim$10 less gamma-ray luminous than V1324~Sco (\citealt{2016ApJ...826..142C}; taking into account the distance estimate for V959~Mon from \citealt{Linford15}).

Another, independent avenue to constrain the shock velocity of novae is their radio light curves. V1324~Sco showed an unusually luminous early radio peak, $L_{\nu, \rm pk} \approx 3\times 10^{30}$ erg s$^{-1}$ at $\nu \approx$ 10 GHz (\S\ref{sec:FirstRadioBump}), compared to $6\times 10^{27}$ erg s$^{-1}$ for V959 Mon and an upper limit of $L_{\nu, \rm pk} \lesssim 2\times 10^{28}$ erg s$^{-1}$ for V339 Del \citep{Chomiuk13, 2014Natur.514..339C, Linford15}.  According to the model presented by \citet{2016MNRAS.463..394V}, the peak synchrotron luminosity is extremely sensitive to the shock velocity, scaling approximately as $L_{\nu, \rm pk} \propto (v_{\rm f}-v_{\rm shell})^{8}$ or steeper (their eq.~48).  Explaining the $\gtrsim$ 1-2 order of magnitude greater synchrotron luminosity of V1324~Sco (compared to V339~Del or V959~Mon) therefore requires a shock velocity which is higher by a factor of $\gtrsim 1.3-1.8$. All else being equal, this velocity difference would alone result in a gamma-ray luminosity $L_{\gamma} \propto (v_{\rm f}-v_{\rm shell})^{2} \gtrsim 2-3$ times higher for V1324 Sco than the other two events.  On the other hand, V1723 Aql and V5589 Sgr showed early radio peaks similar in luminosity to V1324~Sco, but were not detected at all in gamma-ray emission \citep{Krauss11,2016MNRAS.457..887W, 2016MNRAS.460.2687W}.

Clearly, results are mixed as to how the ejecta velocities of V1324~Sco compare to other gamma-ray detected novae. It is unclear how valid a comparison V959~Mon provides, given that ejecta velocities are derived using a wholly different method than in V1324~Sco. We require additional spectroscopic studies of gamma-ray detected novae to provide an appropriate comparison sample with V1324~Sco.  

Another possibility for the high gamma-ray luminosity of V1324~Sco is a particularly high mass loss rate and/or total mass of the fast component, particularly in comparison to the mass in the slow component. From our radio light curve fitting, we do not see an indication that the total mass in V1324~Sco is unusually large. However, from our simple Hubble flow fits, we do not account for multiple components of ejecta and can not make any claims about the relative mass in the fast and slow components.  Because $L_{\gamma} \propto \dot{M}_{\rm f}$, the mass loss rate of the fast component would need only be a factor of $\sim 5$ times higher than in other Fermi-detected novae (for otherwise equal shock velocity) to explain the gamma-ray luminosity of V1324~Sco.

\section{Conclusion}
\label{sec:Conclusion}

We have presented multi-wavelength observations for the most gamma-ray luminous classical nova, V1324 Sco, and demonstrated that this nova was, in all non-gamma-ray observations, a typical classical nova. 
Using the optical photometry and spectra, we classify V1324 Sco as a $D$ (Dusty) photometric class and an \ion{Fe}{2} spectral class nova, both of which are common among classical novae. Ejecta velocities span the range, $1000-2600$ km s$^{-1}$. By fitting the evolution of the thermal radio emission, we derived an ejecta mass a few $\times 10^{-5}$ M$_{\odot}$. This ejecta mass is similar to both theoretical predictions for nova ejecta masses \citep{2005ApJ...623..398Y} and observational determinations of ejecta mass in other novae \citep{Roy12}.

V1324~Sco is the most gamma-ray luminous classical nova discovered to date, but the data we present here do not show anything clearly unusual about this nova. We see strong evidence for shocks, including gamma-ray emission, early time velocity variations in the optical line profiles, and non-thermal radio emission. However, all of these signatures have been seen previously in other novae (although not all together; for example, this is the first time that a double-peaked radio light curve has been observed for a gamma-ray detected nova). 

To explore the shocks and gamma-ray production in V1324~Sco, we present a simple model that invokes ejecta composed of two components: an initial slow component, and a fast secondary component. Using this model, we find that the likely key variables for setting the gamma-ray luminosity of novae are the mass loss rate and velocity of the fast secondary ejecta component. We compare V1324~Sco with two other well-studied gamma-ray detected novae, but do not find clear evidence for higher densities or differential velocities in V1324~Sco. Therefore, the origin of V1324~Sco's high gamma-ray luminosities remains unclear, and can be further explored in the future by comparison with other gamma-ray detected novae.

\section*{Acknowledgments}

We acknowledge and give thanks to the variable star observations from the AAVSO International Database contributed by observers worldwide and used in this research. This research has made use of the AstroBetter blog and wiki. It was funded in part by the \Fermi Guest Investigator grants NNX14AQ36G  (L.\ Chomiuk), NNG16PX24I (C.\ Cheung), and NNX15AU77G and NNX16AR73G (B.\ Metzger). It was also supported by the National Science Foundation (AST-1615084), and the Research Corporation for Science Advancement Scialog Fellows Program (RCSA 23810); S.\ Starrfield gratefully acknowledges NSF and NASA grants to ASU. 

We thank B.\ Broen, B.\ Niedzielski, and R.\ Williams for helpful comments. We are also grateful to anonymous referees for their assistance.

The National Radio Astronomy Observatory is a facility of the National Science Foundation operated under cooperative agreement by Associated Universities, Inc. The LBT is an international collaboration among institutions in the United States, Italy and Germany. LBT Corporation partners are: The University of Arizona on behalf of the Arizona university system; Instituto Nazionale di Astrofisica, Italy; LBT Beteiligungsgesellschaft, Germany, representing the Max-Planck Society, the Astrophysical Institute Potsdam, and Heidelberg University; The Ohio State University, and The Research Corporation, on behalf of The University of Notre Dame, University of Minnesota and University of Virginia. This research is also based on data  collected with the European Southern Observatory telescopes, proposal ID 089.B-0047(I).

\bibliographystyle{apj}

\appendix

\section{A. Radio Light Curve Fitting}
The physical quantities that go into the Hubble flow model (ejecta mass, maximum and minimum ejecta velocities, filling factor, temperature, and distance) are degenerate, and cannot be solved for from a light curve fit alone. To circumvent this issue, we define three composite variables that can be determined uniquely. These three variables that describe the radio light curve are defined as
\begin{equation} \label{eq:Psi}
\Psi \equiv  \frac{T_e v_{\rm max}^2}{D^2}; 
\end{equation}
\begin{equation} \label{eq:Xi}
\Xi \equiv   T_{e}^{-3/2} v_{\rm max}^{-5} \left(\frac{M_{\rm ej}Z}{\mu\, m_H}\right)^{2} g_{ff} f_V^{-1}
\end{equation}
\begin{equation}
\xi \equiv v_{\rm min}/v_{\rm max}.
\end{equation}
Here $T_e$ is the temperature of the emitting region, $v_{\rm max}$ is the maximum velocity of the ejecta, $v_{\rm min}$ is the minimum velocity of the ejecta, $D$ is the distance, $M_{\rm ej}$ is the mass of the ejecta, $Z$ is the average charge of the emitting particles, $f_V$ is the volume filling factor of the ejecta (discussed further in Appendix B), and $\mu\, m_H$ is the average particle mass (we take $\mu$ = 0.6 for an ionized gas of solar abundance). All values are in cgs units. The Gaunt factor $g_{ff}$ at low frequency ($< 10^{12}$ Hz) is $g_{ff} = 9.75 + 0.55\, {\rm ln}\left(T_e^{3/2}/\nu\right)$ \citep{Befeki66}. 

In simple terms, we can think of $\Psi$ as setting the flux scale, as it contains the terms for the blackbody function and the angular size, which combine to give the total flux. We can then think of $\Xi$ as setting the time scale for the ejecta to become optically thin, as $\Xi$ is all of the opacity terms collected into a single variable. 

Writing our expression for total flux density ($S_\nu$) at frequency $\nu$, we find
\begin{equation}
S_\nu  =  \frac{2 k_b}{c^2} \Psi t^{2} \nu^{2} \left[ \int_0^{\xi} a (1-e^{-\tau_1 (a)}) da  + \\
\int_{\xi}^1 a (1-e^{-\tau_2 (a)}) da \right]. 
\end{equation}
The optical depth factors are:
\begin{eqnarray*}
\tau_1 (a) & = & 0.018~\mathrm{ sec}^5 \mathrm{ Hz}^{2} \frac{\Xi~\nu^{-2} t^{-5}}{4\pi[1-\xi]} \int_{\sqrt{\xi - a^2}}^{\sqrt{1- a^2}} \frac{ds}{(a^2 + s^2)^2} ds ; \\
\tau_2 (a) & = & 0.018~\mathrm{ sec}^5 \mathrm{ Hz}^{2} \frac{\Xi~\nu^{-2} t^{-5}}{4\pi[1-\xi ]} \int_0^{\sqrt{1- a^2}} \frac{ds}{(a^2 + s^2)^2} ds.
\end{eqnarray*}
Here, $s$ is the path length through the ejecta, and $a$ is the offset distance between the nova's center and the line of sight (see e.g., \citealt{1979AJ.....84.1619H} for an illustration). Both integrals have had all of their dimensional parameters put into $\Psi$, $\Xi$, and $\xi$ making them unitless. Note that similar composite variables were defined in \citet{1979AJ.....84.1619H}, and the effects of filling factor on the radio light curve were derived by \citet{Heywood04}.

The actual fitting procedure was done using the Markov Chain Monte Carlo program \emph{pymc}~\citep{2010JSS...35..1}. This procedure was selected as it does not enforce a Gaussian distribution of best fit parameters, allowing us to more accurately characterize the full variance of our results. Due to the exceptionally large parameter space occupied by our composite variables (potentially many orders of magnitude), our sampling for the MCMC scheme was done in $\log (\Xi )$ and $\log (\Psi )$ space, and our results are given as such. Our best fit set of parameters are $\log (\Psi ) = -24.487_{-0.031}^{+0.033}$, $\log (\Xi ) = 59.763^{+0.03}_{-0.06}$, and $\xi = 0.447_{-0.079}^{+0.10}$.

\section{B. Deriving the Filling Factor}
\label{sec:FillingFactor}
In this Appendix, we derive a means of determining the filling factor, a parameterization of inhomogeneities (clumpiness) in the ejecta. The following derivation of the filling factor is laid out according the following plan: first we find an analytic expression, in terms of measurables, for the filling factor; then, we detail how we measured the variables and their uncertainty, and then we incorporate the uncertainty into our final calculation. Our method is similar to the one used in~\cite{2006A&A...459..875E}.

\subsection{Derivation}
From~\cite{1998LNP...506..301B}, the filling factor is given by equation 1a,
\begin{equation} \label{eq:fillFactor}
f_V = \frac{\langle n_e^2 \rangle}{n_e^2},
\end{equation}
where $\langle n_e^2 \rangle$ is the average of the density squared. This can be rewritten using equation 4 of the same paper,
\begin{equation} \label{eq:averageSqDensity}
\langle n_e^2 \rangle = \frac{EM}{L},
\end{equation}
where $EM$ is the emission measure and $L$ is the characteristic length of the emitting material. For our purposes we will assume spherical symmetry of the ejecta and say that the characteristic length is $2v_{\rm ej}t$.

We can determine the emission measure from hydrogen recombination lines using equation 3-36 in~\cite{1978ppim.book.....S},
\begin{equation} \label{eq:spitzerEquation}
\int I_\nu d\nu = h\nu \alpha_{mn} \left(\frac{n_p}{n_e} \right) \times 2.46\times 10^{17} E_m.
\end{equation}
where $\alpha_{mn}$ is the effective recombination coefficient for transitions from state $m$ to $n$.
We modify this to be
\begin{equation}
\int I_\nu d\nu = \frac{\int F_\nu d\nu}{\Omega} \approx \frac{F_\nu \Delta \nu}{\Omega} = \frac{F_\lambda \Delta \lambda}{\Omega},
\end{equation}
where $\Omega$ is the solid angle of the source, which we approximate as $(A/D)^2 = \pi (r/D)^2$. Here, $r$ is the ejecta radius, which is just $v_{\rm ej}t$, and $D$ is the distance to the source. 

This leads us to the following expression for emission measure, as determined by measuring the flux in H$\beta$:
\begin{eqnarray}
EM &=& \frac{F_\lambda  \Delta \lambda}{h\nu_{H\beta}\alpha_{42}(2.46\times 10^{17})\Omega}~\mathrm{pc}~ \mathrm{cm}^{-7} \\
&=& \frac{F_\lambda  \Delta \lambda \pi D^2 }{h\nu_{H\beta}\alpha_{42}(2.46\times 10^{17})(v_{\rm ej}t)^2}~\mathrm{pc}~ \mathrm{cm}^{-7}.
\end{eqnarray}
Using this expression for $EM$, we can rewrite equation~\ref{eq:averageSqDensity} as
\begin{equation} 
\langle n_e^2 \rangle = \frac{F_\lambda  \Delta \lambda \pi D^2 }{2 h\nu_{H\beta}\alpha_{42}(2.46\times 10^{17})(v_{\rm ej}t)^3}~\mathrm{pc}~ \mathrm{cm}^{-7}.
\end{equation}
This expression is in terms of pc $\mathrm{cm}^{-7}$, so we must convert it to $\mathrm{cm}^{-6}$. To do this, we multiply by $\left(\frac{3.086\times 10^{18}~\mathrm{cm}}{1~\mathrm{pc}}\right)$, which gives
\begin{equation}
\label{eq:finalAverageSqDensity}
\langle n_e^2 \rangle = \frac{4\pi^2 F_\lambda  \Delta \lambda D^2 }{2 h\nu_{H\beta}\alpha_{42}(v_{\rm ej}t)^3}~\mathrm{cm}^{-6}. 
\end{equation}

Finally, we can determine the density by using spectroscopic line ratios. We will use the [\ion{O}{3}] line ratio to determine density by using equation 5.4 in~\cite{1989agna.book.....O}
\begin{equation}
R_{[O III]} = \frac{j_{\lambda 4959} + j_{\lambda 5007}}{j_{\lambda 4363}} = \frac{7.90 \exp(3.29\times 10^4/T_e)}{1+ 4.5\times 10^{-4}n_e /T_e^{1/2}}.
\end{equation}
This leads to our expression for $n_e$
\begin{equation}
n_e = \frac{T_e^{1/2}}{4.5\times 10^{-4}} \left(\frac{7.90 \exp(3.29\times 10^4/T_e)}{R_{[\textrm{\ion{O}{3}}]}} - 1\right)~\mathrm{cm}^{-3}.
\end{equation}

Squaring the above expression and combining it with equations~\ref{eq:fillFactor} and~\ref{eq:finalAverageSqDensity}, we can now write out our expression for the filling factor.
\begin{equation}
\label{eqn:finalExpression}
f_V =  \left(\frac{2\pi^2 F_\lambda \Delta \lambda D^2 }{h\nu_{H\beta}\alpha_{42}(v_{\rm ej}t)^3}\right)
\times \left[\frac{T_e^{1/2}}{4.5\times 10^{-4}} \left(\frac{7.90 \exp(3.29\times 10^4/T_e)}{R_{[\textrm{\ion{O}{3}}]}} - 1\right)\right]^{-2}. 
\end{equation}

\subsection{Measured Values and Uncertainty}
\label{sec:measuredValues}
The unknown values that we need to solve equation~\ref{eqn:finalExpression} are electron temperature ($T_e$), distance ($D$), ejecta velocity ($v_{\rm max}$), the oxygen line ratio ($R_{[\textrm{\ion{O}{3}}]}$), and the H$_{\beta}$ flux ($F_\lambda \Delta \lambda$). We use the LBT spectrum taken on day $+353$, as it is taken well into the nebular phase and has better spectral response calibration than the SOAR spectrum. Note that the MODS1 instrument was not designed to be a spectrophotometer, and the seeing was twice the width of the slit, so we believe that $\sim 50 \%$ of the flux fell outside of the slit. This issue is negated for line ratios (discussed below), but it does affect absolute line fluxes. Therefore, we will use a fiducial value of $5\%$ for the uncertainty of the line ratios---to account for general calibration uncertainties---and $50 \%$ uncertainty for the absolute line flux.

With this value for the uncertainty, we use the IRAF tool \verb|splot| to measure an  H$_{\beta}$ flux---corrected for the throughput issue mentioned above---of $8.38 \pm 4.19 \times 10^{-15}~\mathrm{ergs}~\mathrm{cm}^{-2}~\mathrm{s}^{-1}$. The line ratio $R_{[\textrm{\ion{O}{3}}]}$ is determined by $j_{\lambda 4959}$, $j_{\lambda 5007}$, and $j_{\lambda 4363}$. We find for these quantities:
\begin{itemize}
\item $j_{\lambda 4959} = 66.2 \pm 3.3 \times 10^{-15}~\mathrm{ergs}~\mathrm{cm}^{-2}~\mathrm{s}^{-1};$
\item $j_{\lambda 5007} = 218.0 \pm 10.9  \times 10^{-15}~\mathrm{ergs}~\mathrm{cm}^{-2}~\mathrm{s}^{-1};$
\item $j_{\lambda 4363} = 4.8 \pm 0.2 \times 10^{-15}~\mathrm{ergs}~\mathrm{cm}^{-2}~\mathrm{s}^{-1}.$
\end{itemize}
As these lines are meant to be a measure of the flux emitted from the source---not the flux measured---we need to make further corrections for interstellar reddening. From~\cite{2015ApJ...809..160F} we know that the reddening is $E(B-V) = 1.16 \pm 0.12$ for V1324 Sco. We use the wavelength specific reddening extinction law of~\cite{1989ApJ...345..245C} (equations 1 and 3), with an $R_V=3.1$, to determine the level of extinction. Doing this, we find reddening corrected fluxes of:
\begin{itemize}
\item $j_{\lambda 4959} = 30.5 \pm 12.4 \times 10^{-13}~\mathrm{ergs}~\mathrm{cm}^{-2}~\mathrm{s}^{-1};$
\item $j_{\lambda 5007} = 95.9 \pm 38.2  \times 10^{-13}~\mathrm{ergs}~\mathrm{cm}^{-2}~\mathrm{s}^{-1};$
\item $j_{\lambda 4363} = 4.5 \pm 2.2 \times 10^{-13}~\mathrm{ergs}~\mathrm{cm}^{-2}~\mathrm{s}^{-1},$
\end{itemize}
and the reddening corrected H$\beta$ line flux is $4.35 \pm 2.84 \times 10^{-13}~\mathrm{ergs}~\mathrm{cm}^{-2}~\mathrm{s}^{-1}$. Note that the uncertainty on the flux values has increase, due to the inclusion of the reddening uncertainty. Using these reddening corrected flux values, we find an $R_{[\textrm{\ion{O}{3}}]}$ value of \OIIIRatio.

For two of the remaining unknown values---$T_e$ and $D$---we use the same values throughout the paper ($T_e = 10^4$ K, $D=$~\AssumedDistance). The remaining value, $v_{\rm ej}$, is derived using the best fit values to the radio data.

\subsection{Final Value}
To determine the final value for $f_v$ we generate distributions of the input variables and plug them into~\ref{eqn:finalExpression}, which gives us a distribution of values for $f_v$. The final value that we quote for $f_v$ is the average of this distribution, and the uncertainty in $f_v$ is the standard deviation of $f_v$. 

We can utilize our distribution of velocities derived in Section~\ref{sec:RadioFitting} to help alleviate some of the uncertainty associated with our measured quantities. From this, and using our canonical nova temperature of $10^4$ K and distance of~\AssumedDistance , we get a filling factor of 
\begin{equation} \label{eq:ff}
f_V = \FillingFactor~.
\end{equation}
The uncertainty is dominated by both the reddening value uncertainty and the fiducial flux calibration uncertainty. Note that filling factor depends on distance as $f_V \propto D^2$, so our lower limit on distance implies that Eq \ref{eq:ff} is also a lower limit on the filling factor.

\section{C. \ion{O}{1} Density Constraints}
\label{sec:DensityLimits}
The strong \ion{O}{1} emission at 7774~\AA~and 8446~\AA~from the Magellan/MIKE spectrum---seen in Figure~\ref{fig:SpectraREDU}---suggests a high density of the emitting material, as the relative strength of the line at 7774~\AA~compared to 8446~\AA~is a measure of the rate of collisional deexcitation~\citep{Williams94, 2012AJ....144...98W}. As 8446~\AA~is a fluorescent line, it should be substantially more dominant than all other \ion{O}{1} lines; the only way for 7774~\AA~to even approach the strength of 8446~\AA~is if there are very high electron densities, such as at a radiative shock front. Note that these \ion{O}{1} lines are originating from the densest portions of the ejecta---perhaps clumps or cold post-shock shells---in contrast with the more ``average'' ejecta densities probed in the previous section on filling factor. 

We can use the oxygen line ratios of $j_{\lambda 7774} /j_{\lambda 6300}$ and $j_{\lambda 8446} /j_{\lambda 6300}$ to place constraints on the temperature, density, and ionizing radiation field of dense, neutral gas present in the nova ejecta (\citealt{1995ApJ...439..346K}; note that the forbidden [\ion{O}{1}] line at 6300~\AA\ can be used as a density diagnostic). This technique works best with high resolution spectra, so we use the MIKE spectrum taken on day $+45$. The other high resolution spectra---taken on day $+3$---had strong, confounding P-Cygni absorption features. After making the necessary reddening corrections (see the above discussion of filling factor for more details) we found an average value of $\log(j_{\lambda 7774} /j_{\lambda 6300}) = 0.39 \pm 0.16$ and an average value of $\log(j_{\lambda 8446} /j_{\lambda 6300}) = 1.08 \pm 0.13$.

We can compare this to the work of \cite{1995ApJ...439..346K} and \cite{1995ApJS...96..325B}, who use a simple model that assumes the rate of excitation can be simply parameterized in terms of the electron number density ($N_e$), temperature ($T_e$), and rate of photoexcitation ($R_p$). If we assume a temperature of $T_e = 10,000$ K, the measured \ion{O}{1} ratios are consistent with density $N_e > 10^{10}$ cm$^{-3}$. Assuming that the density scales as $t^{-3}$, we would expect the density to be a factor of $\sim 10$ times greater on day $+20$ (the time of the first X-ray observation and coincident with gamma-ray detection) than it was on day $+45$. Combined with the fact that we expect the ejecta to have expanded to $\sim$ a few $\times 10^{14}$ cm, we derive a column density $\geq 10^{25}$ cm$^{-3}$ if the filling factor of the \ion{O}{1}-emitting gas is $f_V \approx 0.1$. The filling factor may be a few orders of magnitude smaller for this densest and coldest phase of the nova ejecta, but the column density will remain $>$$\geq 10^{23}$ cm$^{-3}$. This is the column required to absorb X-rays at a few keV \citep{2016arXiv161104532V}.

We can also calculate the total mass that this density implies. Assuming a velocity of $\sim 1,000$ \kms, a mean molecular weight of $2.0 \times 10^{-24}$ grams/particle, and $f_V = 0.1$, this density would correspond to a \ion{O}{1}-emitting ejecta mass of $\approx 2 \times 10^{-4} M_{\odot}$; such a high ejecta mass is at odds with the mass derived from the radio light curve (discussed in Section~\ref{sec:RadioFitting}). This mass could be decreased by a few orders of magnitude if the filling factor of the \ion{O}{1}-emitting ejecta is $<< 0.1$. Another plausible resolution is that the \ion{O}{1}-emitting gas is not evenly distributed as clumps throughout the ejecta, but is instead relegated to  the cooling region behind the radiative shock \citep{2014MNRAS.442..713M}.

\end{document}